\documentclass[draftcls, 12pt,onecolumn,oneside,compress]{IEEEtran}
\usepackage{geometry}
\ifCLASSINFOpdf
\else
   \usepackage[dvips]{graphicx}
\fi
\usepackage{url}

\usepackage{amssymb}
\usepackage{graphicx,color}
\usepackage{mathrsfs}
\usepackage{bm}
\usepackage{amsmath}
\usepackage{amsfonts,amssymb}
\usepackage{extarrows}
\usepackage{graphicx}
\usepackage{algorithm}
\usepackage{algorithmic}
\usepackage{setspace}
\usepackage{array}
\usepackage{epstopdf}
\usepackage{subfigure}
\usepackage{caption}
\usepackage{float}

\usepackage{cite}
\usepackage{url}
\usepackage{stfloats}
\DeclareMathOperator*{\argmax}{arg\,max}
\date{}
\begin{document}

\title{Multi-Camera View Based Proactive BS Selection and Beam Switching for V2X}
\author{Bo Lin, \IEEEmembership{Graduate Student Member, IEEE}, Feifei Gao, \IEEEmembership{Fellow, IEEE}, Yong Zhang, \IEEEmembership{Member, IEEE}, Chengkang Pan, and Guangyi Liu
\thanks{
B. Lin and F. Gao are with the Department of Automation, Tsinghua University, State Key Lab of Intelligent Technologies and Systems, Tsinghua University, State Key for Information Science and Technology (TNList), Beijing 100084, P. R. China (e-mail: feifeigao@ieee.org; linb20@mails.tsinghua.edu.cn).

Y. Zhang is with the Beijing Key Laboratory of Multimedia and Intelligent Software Technology, Beijing Institute of Artificial Intelligence, Faculty of Information Technology, Beijing University of Technology, Beijing 100124, P. R. China (e-mail: zhangyong2010@bjut.edu.cn).

C. Pan and G. Liu are with the China mobile communication research Institute, Beijing 100053, P. R. China (e-mail: liuguangyi@chinamobile.com; panchengkang@chinamobile.com).
}\vspace{-15mm}}
\maketitle
\thispagestyle{empty}
\begin{abstract}
Due to the short wavelength and large attenuation of millimeter-wave (mmWave), mmWave BSs are densely distributed and require beamforming with high directivity.
When the user {moves} out of the coverage of the current BS or is severely blocked, the mmWave BS must be switched to ensure the communication quality.
In this paper, we proposed a multi-camera view based proactive BS selection and beam switching that can predict the optimal BS of the user in the future frame and switch the corresponding beam pair.
Specifically, we extract the features of multi-camera view images and a small part of channel state information (CSI) in historical frames, and dynamically adjust the weight of each modality feature.
Then we design a multi-task learning module to guide the network to better understand the main task, thereby enhancing the accuracy and the robustness of BS selection and beam switching.
Using the outputs of all tasks, a prior knowledge based fine tuning network is designed to further increase the BS switching accuracy.
After the optimal BS is obtained, a beam pair switching network is proposed to directly predict the optimal beam pair of the corresponding BS.
Simulation results in an outdoor intersection environment show the superior performance of our proposed solution under several metrics such as predicting accuracy, achievable rate, harmonic mean of precision and recall.
\end{abstract}


\IEEEpeerreviewmaketitle

\section{Introduction}\label{s_introduction}
In 5G network, millimeter wave (mmWave) frequency communication \cite{agiwal2016next} is used to provide extremely high data rates in Gbps class.
Despite the high transmission rate, mmWave communication have short wavelengths, and are seriously affected by atmospheric loss and rainfall attenuation, resulting in short transmission distances \cite{li2016path}.

In order to {compensate} for the large path loss of mmWave propagation, beamforming is naturally used in mmWave communication systems {with} multiple-input and multiple-output (MIMO) technique  \cite{roh2014millimeter}.
Beamforming concentrates the signal energy in the direction with the highest signal-to-noise ratio (SNR) for data transmission and maximizes the transmission rate.
However, due to weak diffraction ability, narrow beam characteristic, and vehicles' mobility, mmWave beams are sensitive to blockage\cite{niu2015survey}.
Field measurements have shown that blockages caused by large vehicles on the road can attenuate the received signal power by more than 20 dB \cite{boban2019multi}.
Such drastic power degradation can lead to sudden link outages and massive data loss.

Coordinated multipoint (CoMP), could release the outage caused by blockage since it can combine multi-base stations to cooperatively serve mobile users \cite{maccartney2017base,maamari2016coverage}.
In \cite{maamari2016coverage}, the authors demonstrated that in a stochastic geometry framework, co-operation from randomly located base stations decreases the probability of outage and increases the coverage probability.
However, the premise of CoMP communication is that the user obtains precise channel state information (CSI) between him and multiple base stations (BSs).
Since beamforming in mmWave communications is based on MIMO arrays, the grow in the number of antennas will lead to a dramatic increase in the pilot overhead \cite{larsson2014massive}.
Moreover, simultaneously estimating the CSI between the user and multiple BSs will further increase the pilot overhead several times.
Another method to address interruptions caused by blockage is BS handoff.
When the communication between the BS and the user is interrupted, another BS will be switched to communicate with the user.
The conventional handoff methods include horizontal handoff and vertical handoff \cite{nasser2006handoffs}.
Horizontal handoff changes the BS to a geographically neighboring BS supporting the same network technology.
However, in mmWave systems, the BSs are more densely distributed, and there may be multiple BSs at the same distance from the user.
It is then difficult to select the optimal handoff target only by geographical location.
Vertical handoff is between two BSs that have different wireless access technologies.
Both horizontal and vertical handoff use wireless communication indicators such as received signal strength (RSS) to determine whether handoff is required.
However, the judgment of handover is not preemptive, and the communication metrics are already degraded before handoff.

In recent years, artificial intelligence has been unstoppably used in physical layer wireless communication.
Channel estimation, beamforming, signal detection, modulation and demodulation, etc., are widely combined with deep learning, bringing a new development attitude to physical layer communication \cite{8795533,9175003,lin2021deep,ye2017power,xia2019deep,gao2022data,9791486,bu2020adversarial}.
In \cite{alkhateeb2018machine}, the BS learns how to predict the blockage using their observations of adopted beamforming vectors.
Since the power of the high-frequency mmWave received signal principally comes from the direct diameter, the blockages of the mmWwave signals are mainly caused by the occlusion of the line-of-sight (LOS).
Hence, image captured by  the BS is used for blockage prediction as a new dimension of information.
In \cite{charan2021vision}, Charan et al. proposed a vision-aided solution to predict blockage and user handoff.
However, when the user is occluded, a single perspective will not be able to continuously track the user, and the user will be disconnected.
Moreover, such solution predicts whether blockage is imminent, and simply judge whether to handoff according to the blockage situation.
In \cite{8941039}, the authors proposed a reinforcement learning based proactive framework for leveraging time consecutive camera images in handover decision problems.
However, only two BSs are used for wireless communication.
Once the handoff is predicted, the user will directly switch to another BS without evaluation of such BS.
How to choose the optimal BS to switch has not been considered yet.

In this paper, we propose a multi-camera view based proactive BS selection network (PBSN) in a multi-BS system.
We equip each BS with an RGB camera to continuously capture scene vision.
The images from multiple cameras work together to predict the optimal BS at the next moment.
The proposed PBSN consists of three parts, namely multi-modal feature extraction module (FEM), multi-task learning \cite{zhang2021survey} based BS selection module (BSM), and prior knowledge based fine tuning module (FTM).
The FEM extracts the geometric features from multi-camera view images, and electromagnetic features from partial channel.
The BSM chooses the prediction of the user's future area and the prediction of the future blockage between the user and each BS as the two subtasks.
The sub-tasks and the main task share the shallow network, which enables the network to \textbf{better understand the main task of selecting the BS}.
In the FTM, we consider the effect of blockage on BS selection, and design a tuning module to \textbf{further increase the BS selection accuracy}.
After selecting the optimal BS, we design a multi-task learning based beam pair switching network (BPSN).
The BPSN predicts the optimal beam pair that maximizes the transmission rate according to the images of the present frame, the user location and the partial channel of the BS.
Since the switching of beam pair is strongly correlated with the channel matrix, we choose the channel reconstruction task as a sub-task of switching the beam pair.

The remainder of this paper is organized as follows.
Section \ref{s_system_and_channel_model} introduces the channel model and system model.
Section \ref{s_predict1} designs the framework of the multi-camera view based proactive BS selection network.
Section \ref{s_predict2} presents the image based beam pair switching network.
Section \ref{simulation}  provides the simulation results and Section \ref{conclusion} draws the conclusion.

\section{Channel Model and System Model}\label{s_system_and_channel_model}

\subsection{Channel Model}\label{s_channel_model}
We adopt a 3-D geometric based channel model \cite{heath2016overview} where signal emitted by the transmitter reaches the receiver from multiple paths through reflection, diffraction, and refraction \cite{sayeed2010wireless}.
Denote $\alpha_{l}$ as the attenuation coefficient of the $l$-th path, $\phi_{l}^{a,D}$ as the azimuth angle of departure (AoD), $\phi_{l}^{e,D}$ as the elevation AoD , $\phi_{l}^{a,A}$ as the azimuth angle of arrival (AoA), $\phi_{l}^{e,A}$ as the elevation AoA, $\vartheta_{l}$ as the phase, and $\tau_{l}$ as the propagation delay.
The channel matrix $\mathbf{H}$ is given by \cite{sayeed2002deconstructing}
\begin{equation}\label{channelmodel}
\mathbf{H} = \sum_{l=1}^{L}\alpha_{l}e^{j\vartheta_{l}+2\pi \tau_{l}B}\mathbf{a}(\phi_{l}^{a,A},\phi_{l}^{e,A})\mathbf{a}^{*}(\phi_{l}^{a,D},\phi_{l}^{e,D}),
\end{equation}
where $B$ is the signal bandwidth, and $\mathbf{a}(\phi_{l}^{a,A},\phi_{l}^{e,A})$ and $\mathbf{a}(\phi_{l}^{a,D},\phi_{l}^{e,D})$ are the steering vectors at the arrival and departure sides.
The mathematical expression of $\mathbf{a}(\cdot)$ is
\begin{equation}\label{4}
\mathbf{a}(\phi_{l}^{a,A},\phi_{l}^{e,A}) = \mathbf{a}_{z}(\phi_{l}^{e,A})\otimes \mathbf{a}_{y}(\phi_{l}^{a,A},\phi_{l}^{e,A})\otimes \mathbf{a}_{x}(\phi_{l}^{a,A},\phi_{l}^{e,A}),
\end{equation}
where $\mathbf{a}_{x}(\cdot)$, $\mathbf{a}_{y}(\cdot)$, $\mathbf{a}_{z}(\cdot)$ are the BS array response vectors in the $x$, $y$, and $z$ directions (the operation is the same for the AoD).
The operators $\mathbf{a}_{x}(\cdot)$, $\mathbf{a}_{y}(\cdot)$, $\mathbf{a}_{z}(\cdot)$ are defined as
\begin{equation}
\begin{aligned}
\mathbf{a}_{x}(\phi_{l}^{a,A},\phi_{l}^{e,A}) &= [1,e^{j\frac{d_{x}}{\lambda}sin(\phi_{l}^{e,A})cos(\phi_{l}^{a,A})},\cdots,e^{j\frac{d_{x}}{\lambda}(N_{x}-1)sin(\phi_{l}^{e,A})cos(\phi_{l}^{a,A})}]\\
\mathbf{a}_{y}(\phi_{l}^{a,A},\phi_{l}^{e,A}) &= [1,e^{j\frac{d_{y}}{\lambda}sin(\phi_{l}^{e,A})sin(\phi_{l}^{a,A})},\cdots,e^{j\frac{d_{y}}{\lambda}(N_{y}-1)sin(\phi_{l}^{e,A})sin(\phi_{l}^{a,A})}]\\
\mathbf{a}_{z}(\phi_{l}^{e,A}) &= [1,e^{j\frac{d_{z}}{\lambda}cos(\phi_{l}^{e,A})},\cdots,e^{j\frac{d_{z}}{\lambda}(N_{z}-1)cos(\phi_{l}^{e,A})}],
\end{aligned}
\end{equation}
where $\lambda$ is the carrier wavelength, while $d_{x}$, $d_{y}$, $d_{z}$ are the antenna spacings in the $x$-, $y$-, and $z$- direction.

\begin{figure}[t]
\centering
\includegraphics[width=0.5\textwidth]{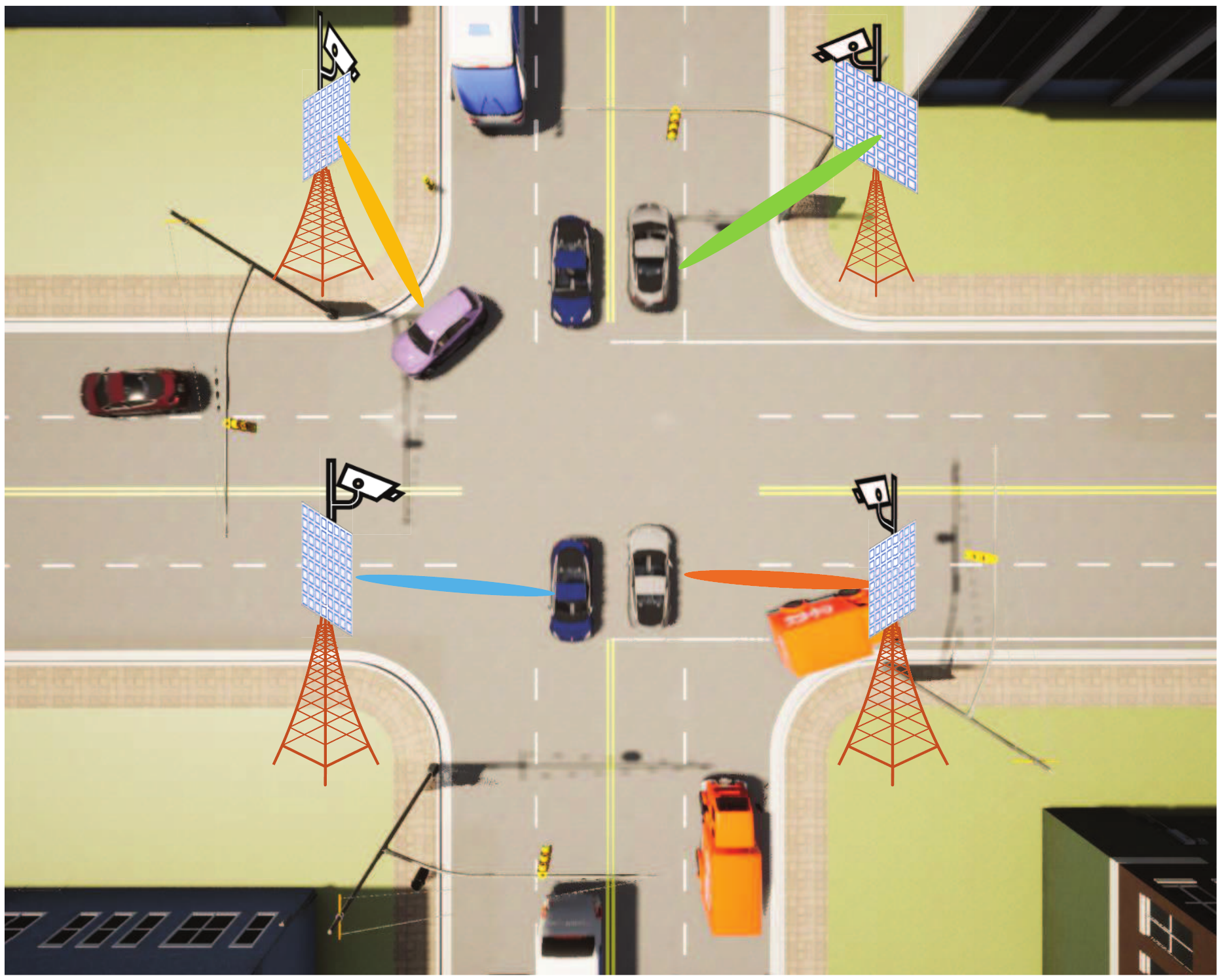}
\caption{The multi-BS system with cameras.}
\label{crossroad}
\end{figure}

\subsection{System Model}\label{system}
We consider a communication scenario at an crossroad as shown in Fig. \ref{crossroad}, where the service users are the vehicles.
Each corner of the crossroad is equipped with a BS and a camera.
The visual information of the camera is used as sensing information to assist wireless communication.

Each BS is equipped with $N_t$ antennas, and each user is equipped with $N_r$ antennas.
Define $\mathbf{H}^{u,b}_{i} \in \mathbb{C}^{N_r \times N_t}$ as the channel matrix from the $b$-th ($b \in \mathbf{B}=\{1,2,3,4\}$) BS to the $u$-th user at the $i$-th frame.
Assume that both the BS and the user have one radio frequency chain. Analog beamforming is performed at both the transmitter and the receiver.
Define $\bm{f}_{t,i}$ as the transmit beamforming vector at BS, and $\bm{f}_{r,i}$ as the receive beamforming vector at the user.
Define the transmit signal as $s$.
Then the receive signal is
\begin{equation}\label{signal_transmission}
\begin{aligned}
y = \bm{f}_{r,i}^{\rm{H}}\mathbf{H}^{u,b}_{i}\bm{f}_{t,i}s + \bm{f}_{r,i}^{\rm{H}}\bm{n},
\end{aligned}
\end{equation}
where $\bm n \in \mathcal{CN}(0,\sigma^{2}\mathbf{I})$ is the Gaussian noise.
The transmit beamforming vector $\bm f_{t,i}$ is chosen from the transmit beam codebook $\mathcal{F}_{t}=\{\bm f_{t,1},\bm f_{t,2},\dots,\bm f_{t,N_t}\}$, while the receive beamforming vector $\bm f_{r,i}$ is chosen from the receive beam codebook $\mathcal{F}_{r}=\{\bm f_{r,1},\bm f_{r,2},\dots,\bm f_{r,N_r}\}$.
The best BS for communication and the corresponding pair of beamforming vectors $(\bm f_{t,i},\bm f_{r,i})$ are $(b^{opt}, \bm f_{t,i}^{opt},\bm f_{r,i}^{opt})$ that maximizes transmission rate
\begin{equation}\label{rate}
\begin{aligned}
(b^{opt}, \bm f_{t,i}^{opt},\bm f_{r,i}^{opt}) = \argmax_{b \in \mathbf{B},\,\bm{f}_{r,i} \in {\mathcal{F}_{r}},\,\bm{f}_{t,i} \in {\mathcal{F}_{t}}} \text{log}_{2}\left(1+\text{SNR}\left|\bm{f}_{r,i}^{\rm{H}}\mathbf{H}^{u,b}_{i}\bm{f}_{t,i}\right|^{2}\right),
\end{aligned}
\end{equation}

Define the RGB images captured by the four cameras at the $i$-th frame as $\mathbf I_{i}^1$, $\mathbf I_{i}^2$, $\mathbf I_{i}^3$ and $\mathbf I_{i}^4$.
From the image captured by any BS, we can obtain the 2D information of the scene, which is called scene semantics \cite{weng2021semantic}.
However, from the images of multiple views, the complete 3D information of the scene can be reproduced \cite{kim2009multi}.
Let us use $\mathcal{S}_i$ to represent the scene information of the $i$-th frame, including the users' locations, shapes, sizes, relative occlusions between each other, and so on.
These scene information determines the parameters in equation (\ref{channelmodel}).
According to the channel model in the Section \ref{s_channel_model}, the channel matrix can be theoretically calculated.
Hence, we can obtain the optimal BS and beam pair at the $i$-th frame according to equation (\ref{rate}).
In other word, there is a certain mapping relationship between the images $\mathbf I_{i}^1$, $\mathbf I_{i}^2$, $\mathbf I_{i}^3$, $\mathbf I_{i}^4$ and $(b^{opt}, \bm f_{t,i}^{opt},\bm f_{r,i}^{opt})$.
\begin{equation}\label{mapping}
\bm{\Phi}_1:\left\{\mathbf I_{i}^1, \mathbf I_{i}^2, \mathbf I_{i}^3, \mathbf I_{i}^4\right\} \rightarrow \mathcal{S}_i \rightarrow \left\{(b^{opt}, \bm f_{t,i}^{opt},\bm f_{r,i}^{opt})\right\},
\end{equation}

In order to ensure the smoothness of the wireless link, we hope that the system can accurately predict the optimal BS and beam pair in the future frame. 
In the real world, it is impossible for the state of an object to mutate.
Therefore, the future state of the scene can be predicted by analyzing the state at the historical moment \cite{liu1998mobility}.
For example, the scene information $\mathcal{S}_i$ can be predicted by images of historical $T$ frames $\left\{\mathbf I_{i}^1, \cdots, \mathbf I_{i+T-1}^1, \mathbf I_{i}^2, \cdots, \mathbf I_{i+T-1}^2, \mathbf I_{i}^3, \cdots, \mathbf I_{i+T-1}^3, \mathbf I_{i}^4, \cdots, \mathbf I_{i+T-1}^4\right\}$.
\begin{equation}\label{mapping2}
\begin{aligned}
\bm{\Phi}_2:\left\{\mathbf I_{i}^1, \cdots, \mathbf I_{i+T-1}^1, \mathbf I_{i}^2, \cdots, \mathbf I_{i+T-1}^2, \mathbf I_{i}^3, \cdots,\mathbf I_{i+T-1}^3, \mathbf I_{i}^4, \cdots, \mathbf I_{i+T-1}^4\right\} \rightarrow \mathcal{S}_i,
\end{aligned}
\end{equation}
Through (\ref{mapping}) and (\ref{mapping2}), we know that the optimal BS and beam pair in the future time step can be predicted by the historical consecutive images from multiple views
\begin{equation}\label{mapping3}
\begin{aligned}
\bm{\Phi}:\left\{\mathbf I_{i}^1, \cdots, \mathbf I_{i+T-1}^1, \mathbf I_{i}^2, \cdots, \mathbf I_{i+T-1}^2, \mathbf I_{i}^3, \cdots,\mathbf I_{i+T-1}^3, \mathbf I_{i}^4, \cdots, \mathbf I_{i+T-1}^4\right\} \rightarrow \left\{(b^{opt}, \bm f_{t,i+T}^{opt},\bm f_{r,i+T}^{opt})\right\}.
\end{aligned}
\end{equation}

As the exact mathematical function of mapping (\ref{mapping3}) is hardly to obtain, we adopt deep neural networks (DNN) to fit $\bm\Phi$ with the aided of training data.
Then, the prediction function can be described as
\begin{equation}\label{extrapolation_function}
\left\{(b^{opt}, \bm f_{t,i+T}^{opt},\bm f_{r,i+T}^{opt})\right\} = f(\left\{\mathbf I_{i}^1, \cdots, \mathbf I_{i+T-1}^4\right\},\bm{\Theta_{e}}),
\end{equation}
where $\bm{\Theta_{e}}$ is the parameters of DNN.
For (\ref{extrapolation_function}), there are $N_{bs}\times N_t\times N_r$ possible outputs, which makes the prediction difficult.
To reduce the difficulty of the prediction, we split (\ref{extrapolation_function}) into two functions (two tasks):
(i) Predict the optimal BS based on images captured by multiple BSs (Section \ref{s_predict1});
(ii) Forecast the optimal beam pair under the predicted BS (Section \ref{s_predict2}).

\section{Multi-Camera View Based Proactive BS Selection}\label{s_predict1}
We here propose a multi-camera view based proactive BS selection network (PBSN) that contains three modules: multi-modal feature extraction module (FEM), multi-task learning based BS selection module (BSM), and prior knowledge based fine tuning module (FTM).
The framework of PBSN is shown in Fig. \ref{system}.

\begin{figure}[t]
\centering
\includegraphics[width=0.8\textwidth]{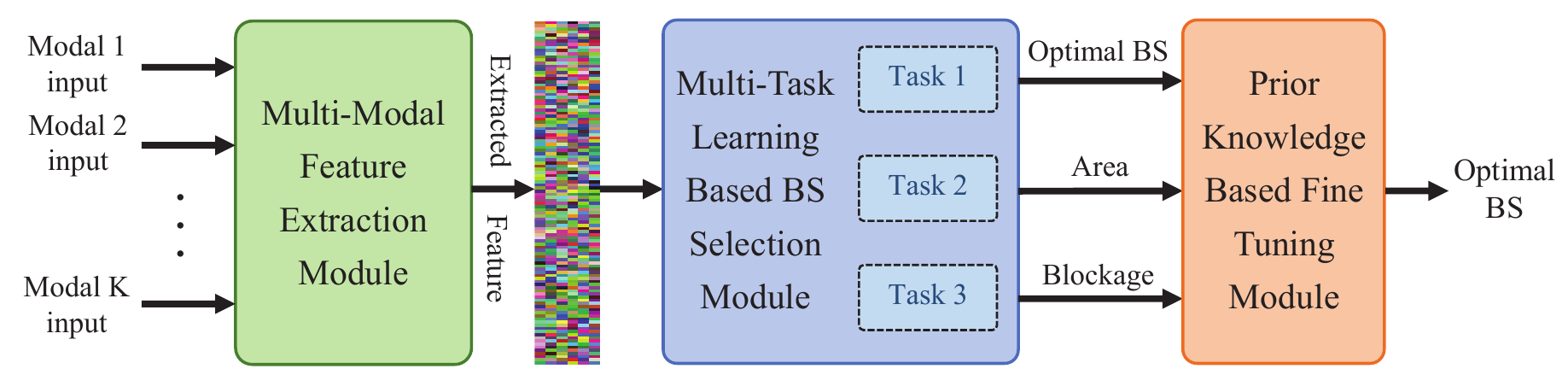}
\caption{The framework of PBSN.}
\label{system}
\end{figure}
\subsection{Multi-Modal Feature Extraction Module}\label{s_multi_modal}
\begin{figure}[t]
\centering 
\includegraphics[width=0.57\textwidth]{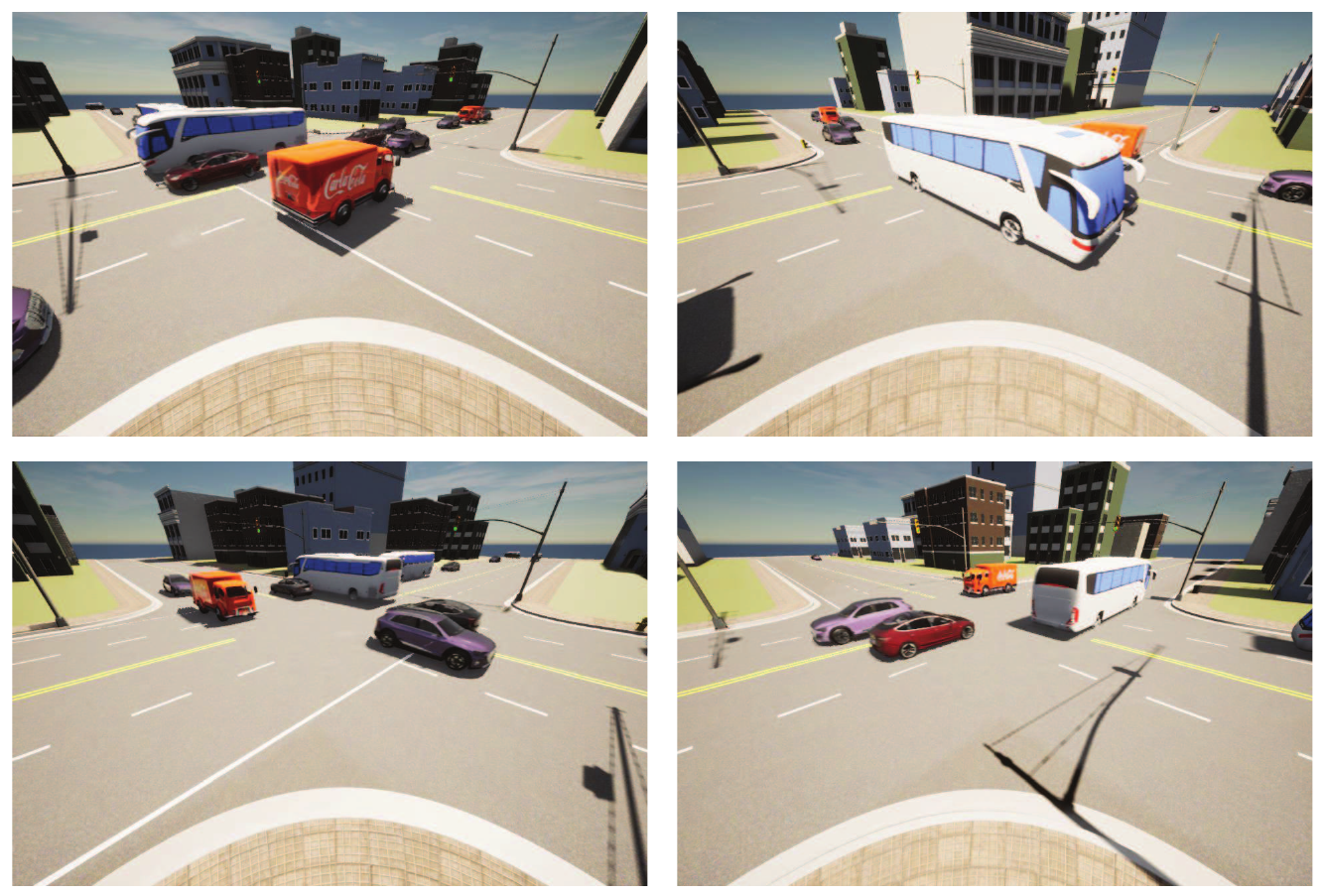}
\caption{The crossroad captured by cameras from four views.}
\label{four_img_detection}   
\end{figure}

Although images from multiple views contain 3D information of the scene, it is still difficult to predict the optimal BS from only four images.
There are two main reasons why multi-modal input needs to be introduced.
\\1) Fig. \ref{four_img_detection} shows images of the crossroad captured by cameras from four views.
Using the image processing and deep learning based object detection algorithm, all the objects in the images can be detected.
Combining the bounding boxes in multiple views, we can calculate the 3D locations of all the objects in the scene.
However, a critical problem is that the neural network cannot know which user the BS is serving.
Hence, we need a extra modal input for user identification.
In this paper, we assume that the BS knows which user is communicating with it.
We need to find a typical parameter that is unique to each user and feed it into the neural network as the user identification.
In real scenarios, the positions of different users cannot overlap at the same time.
Therefore, the \textbf{user's location} is an effective identifier of the user and can be used as an additional modality.
\\2) As described in Section \ref{s_system_and_channel_model}, the optimal BS and beam pair are selected according to the channel matrix.
The parameters that constitute the channel matrix are $\alpha_{l}$, $\phi_{l}^{a,D}$, $\phi_{l}^{e,D}$, $\phi_{l}^{a,A}$, $\phi_{l}^{e,A}$, $\vartheta_{l}$ and $\tau_{l}$.
These parameters are affected by the geometric and electromagnetic properties of the scene.
The geometric features of the scene can be extracted from multi-view images.
However, the electromagnetic properties of the scene are complexly related to many factors such as building materials, vehicle surface materials, and weather conditions.
In \cite{gao2021fusionnet} it is demonstrated that a part of channel matrices contains the electromagnetic features of the environment and is helpful for perception.
Therefore, a part of the \textbf{user's channel} is chosen as an extra modality for better awareness of the electromagnetic environment.

Next, we introduce how to acquire the information of the three modalities.
The images are captured by cameras set up on the BS.
The partial channel matrix can be obtained by sending pilots on partial antennas and performing channel estimation.
Then we introduce how to acquire the user's location.
In recent years, there have been many excellent target detection algorithms in the field of image processing, such as r-cnn \cite{girshick2015fast} and YOLO \cite{redmon2016you}.
Object detection algorithms can identify potential users from an image and mark them with bounding boxes.
We obtain the 2D position of the user in the image of one certain view. 
The 3D position can be calculated by combining the 2D positions from different views.
Although the user may be occluded by large vehicles, as long as there is more than one viewpoint to see the user, we can calculate the 3D position of the user, whose steps can be referred to in Appendix \ref{appendix}.
So far, we have obtained the multi-camera view images, the user's location, and the user's partial channel.

\begin{figure}[t]
\centering
\includegraphics[width=0.65\textwidth]{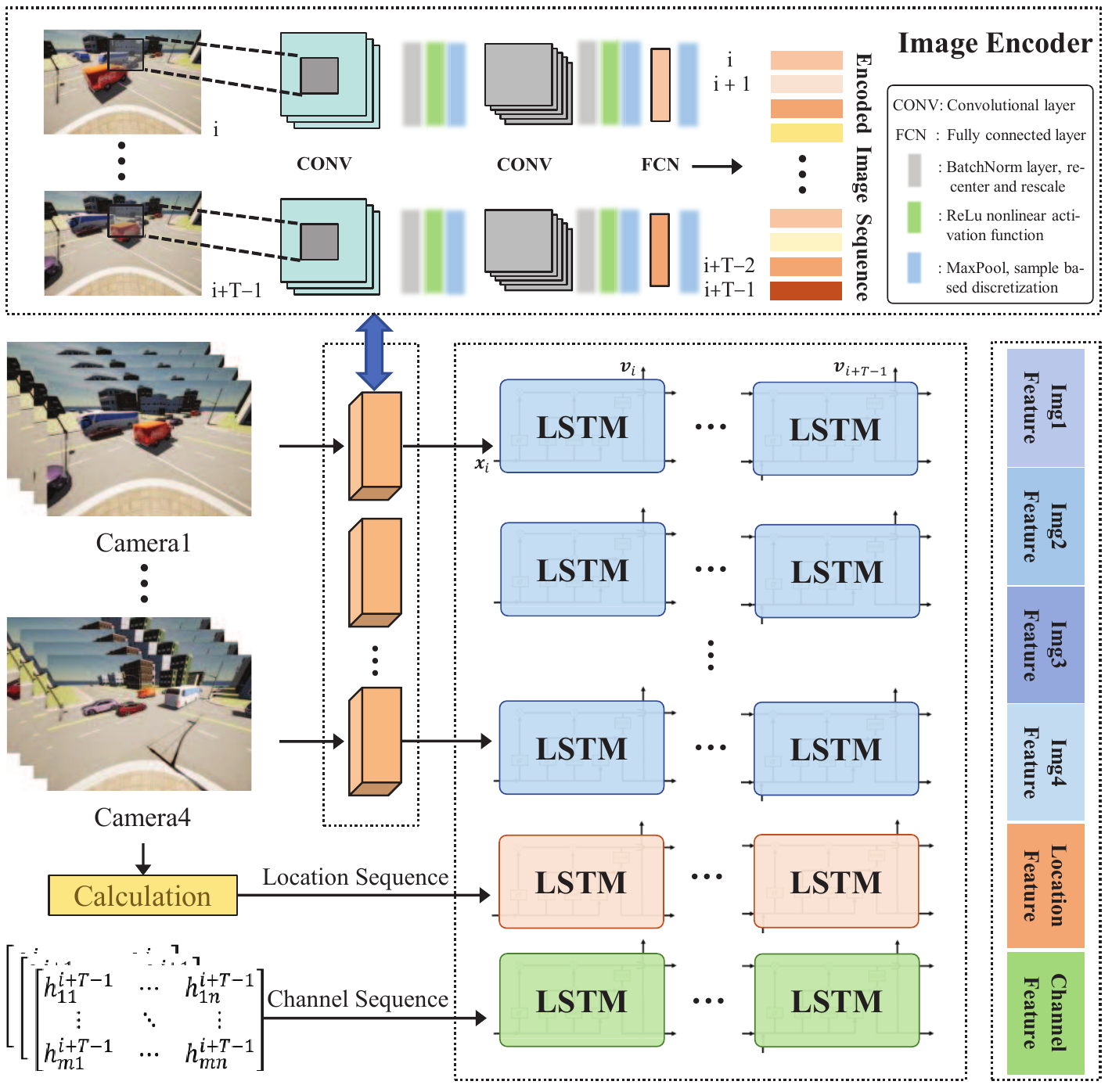}
\caption{FEM: multi-modal feature extraction module.}
\label{FEM}
\end{figure}

After obtaining the three modalities in each frame, we integrate information from consecutive $T$ frames for subsequent feature extraction.
The location sequence ($\mathbf{LS}_i$) of the user is defined as
\begin{equation}\label{location_feature_sequence}
    \mathbf{LS}_{i} = \left[\mathbf{Loc}_{i},\mathbf{Loc}_{i+1},\cdots,\mathbf{Loc}_{i+T-1}\right],
\end{equation}
where $\mathbf{Loc}_{i}\in \mathbb{R}^{2}$ is the 2D coordinates user's location in the $i$-th frame.
The channel sequence ($\mathbf{CS}_i$) of the user is denoted as
\begin{equation}\label{channel_matrix_sequence}
    \mathbf{CS}_{i} = \left[\mathbf{H}_{i,s},\mathbf{H}_{i+1,s},\cdots,\mathbf{H}_{i+T-1,s}\right],
\end{equation}
where $\mathbf{H}_{i,s}$ is the down-sampled channel matrix in the $i$-th frame.

For the visual modality, since an image contains too much redundant information, we first encode it by an image encoder.
The image encoder consists of several convolutional neural networks (CNNs) \cite{krizhevsky2012imagenet}, that are widely used in various image processing projects.
Each CNN includes several convolutional layers and a linear layer.
The mapping function of each convolutional layer can be defined as $\textup{CNNCell}(\cdot)$
\begin{equation}\label{CNN_layer}
\begin{aligned}
\textup{CNNCell}(\mathbf x) = \textup{MaxPool}(\textup{ReLu}(\textup{BatchNorm}(\textup{Conv}(\mathbf x)))),
\end{aligned}
\end{equation}
where $\textup{Conv}(\cdot)$ is the function of the convolutional neural layer, $\textup{BatchNorm}(\cdot)$ is the function that normalizes the layers' inputs by re-centering and re-scaling, $\textup{ReLu}(\cdot)$ is the nonlinear activation function, $\textup{MaxPool}(\cdot)$ is a sample-based discretization function to down-sample an input representation.
The mapping function of the linear layer is defined as $\omega(\cdot)$
\begin{equation}\label{linear_layer}
\begin{aligned}
\omega(\mathbf x) = \textup{ReLu}(\mathbf{W}*\mathbf x+\mathbf{bias}),
\end{aligned}
\end{equation}
where $\mathbf{W}$ is the weight matrix of the linear layer, $\mathbf{bias}$ is the corresponding bias vector.
The output of the image encoder of the $b$-th camera is denoted as $\mathbf{IE}_{i}^{b}$
\begin{equation}\label{image_feature}
\begin{aligned}
\mathbf{IE}_{i}^{b} &= \omega\left(\textup{CNNCell}_{Mi_{n}}^{b}(\cdots \textup{CNNCell}_{Mi_{1}}(\mathbf I_{i}^b))\right).
\end{aligned}
\end{equation}
Then, the encoded image sequence of the $b$-th camera is denoted $\mathbf{IS}^{b}_{i}$
\begin{equation}\label{image_feature_sequence}
    \mathbf{IS}^{b}_{i} = \left[\mathbf{IE}_{i}^{b},\mathbf{IE}_{i+1}^{b},\cdots,\mathbf{IE}_{i+T-1}^{b}\right].
\end{equation}

\begin{figure}[t]
\centering
\includegraphics[width=0.5\textwidth]{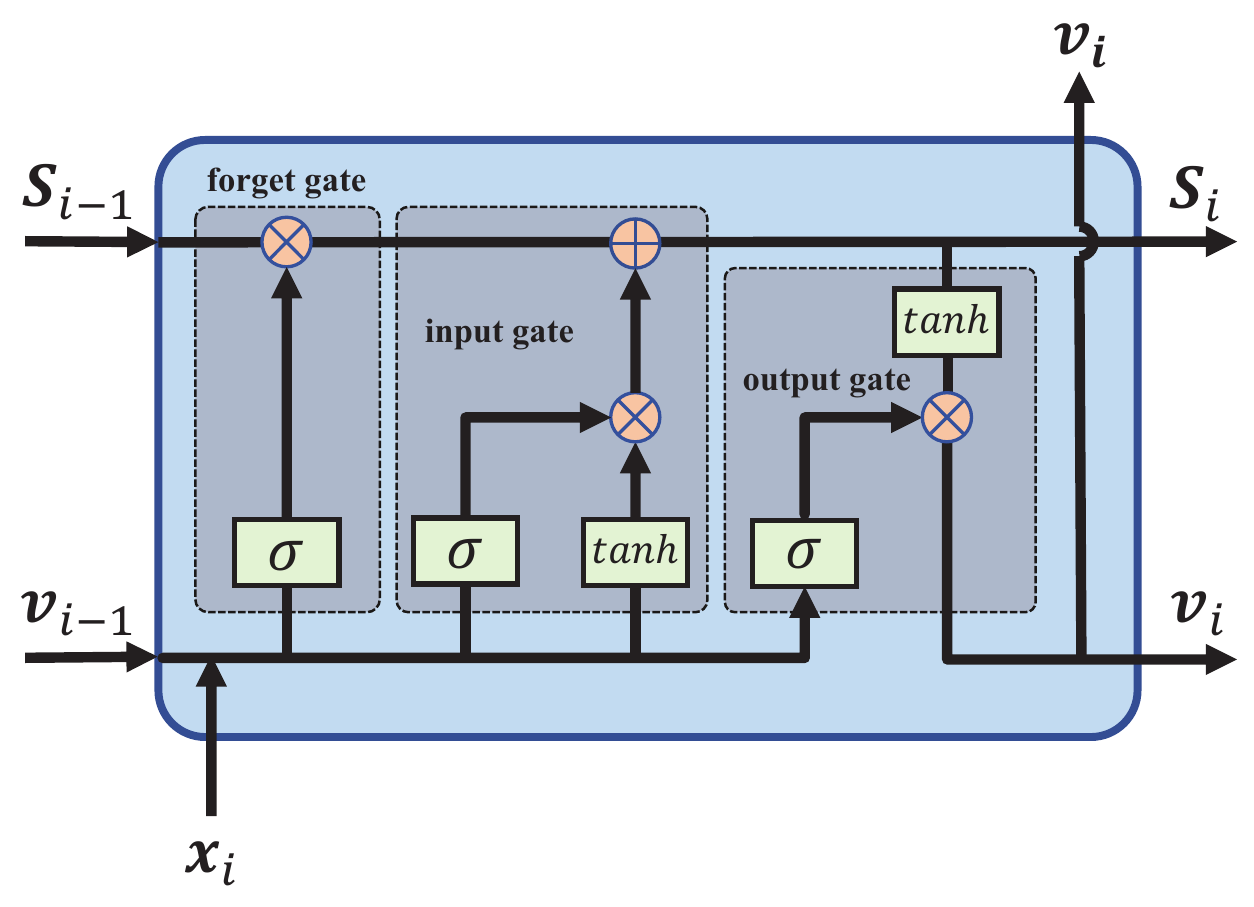}
\caption{The structure of LSTM.}
\label{LSTM}
\end{figure}

We design the FEM to extract the geometric and electromagnetic features of the scene as well as the user's identification features from $\mathbf{LS}_i$, $\mathbf{CS}_i$, and $\mathbf{IS}^b_i$ ($b=1,2,3,4$).
We adopt Long Short Term Memory (LSTM) network \cite{hochreiter1997long}, a typical network in processing time-ordered sequences, to extract the features of the three modalities in $T$ frames.
As shown in Fig. \ref{LSTM}, the $T$ frame LSTM consists of $T$ cells, whose parameters include $\bm x_i$, $\bm v_i$ and $\bm{S}_i$, where $\bm x_i$ is the input of the $i$-th frame, $\bm v_i$ is the output, and $\bm{S}_i$ is the cell state.
In the $i$-th frame, the network decides to discard some of the past cell state through the forget gate.
The output of the forget gate is devoted as $\bm f_{i}$
\begin{equation}\label{forget_gate}
\bm f_{i} = \sigma\left(\bm W_f\cdot\left[\bm v_{i-1},\bm x_i\right]+\bm b_f\right),
\end{equation}
where $\sigma(\cdot)$ is the sigmoid function
\begin{equation}\label{sigmoid_function}
\sigma(x) = \frac{1}{1+e^{-x}}\in\left(0,1\right),
\end{equation}
while $\bm W_f$ and $\bm b_f$ are the weight matrix and the bias vector of the forget gate.
The cell state retained by the previous cell is $\bm f_{i}\odot \bm{S}_{i-1}$.
The input cell state of the $i$-th frame is defined as $\tilde{\bm S}_{i}$
\begin{equation}\label{state_hat_i}
\tilde{\bm S}_{i} = tanh\left(\bm W_c\cdot \left[\bm v_{i-1},\bm x_i\right]+\bm b_c\right).
\end{equation}
Then we use the input gate to decide what information in $\tilde{\bm S}_{i}$ can be saved into the current cell state $\bm S_i$.
The input gate generates a vector to control which information in $\tilde{\bm S}_{i}$ will be reserved in the current state.
The output of the input gate is denoted as $\bm s_{i}$
\begin{equation}\label{forget_gate}
\bm s_{i} = \sigma\left(\bm W_s\cdot\left[\bm v_{i-1},\bm x_i\right]+\bm b_s\right).
\end{equation}
Then we combine the input cell state with the past state in memory to obtain an updated cell state $\bm S_i$ as
\begin{equation}\label{state_i}
\bm{S}_{i} = \bm f_i \odot \bm{S}_{i-1} + \bm s_{i}\odot\tilde{\bm S}_{i},
\end{equation}
We use a sigmoid layer to determine which part of the cell state will output
\begin{equation}\label{output_gate}
\bm o_{i} = \sigma\left(\bm W_o\cdot\left[\bm v_{i-1},\bm x_i\right]+\bm b_o\right),
\end{equation}
and then the output is
\begin{equation}\label{output}
\bm v_i = \bm o_{i}\cdot \textup{tanh}(\bm{S}_i).
\end{equation}

Combining the output of LSTM cell for each frame, the extracted feature is
\begin{equation}\label{extracted_feature}
\bm F_i = \left[\bm v_i^T, \bm v_{i+1}^T, \cdots, \bm v_{i+T-1}^T\bm \right]^T.
\end{equation}

\begin{figure}
  \begin{minipage}[t]{0.5\linewidth}
    \centering
    \includegraphics[scale=0.8]{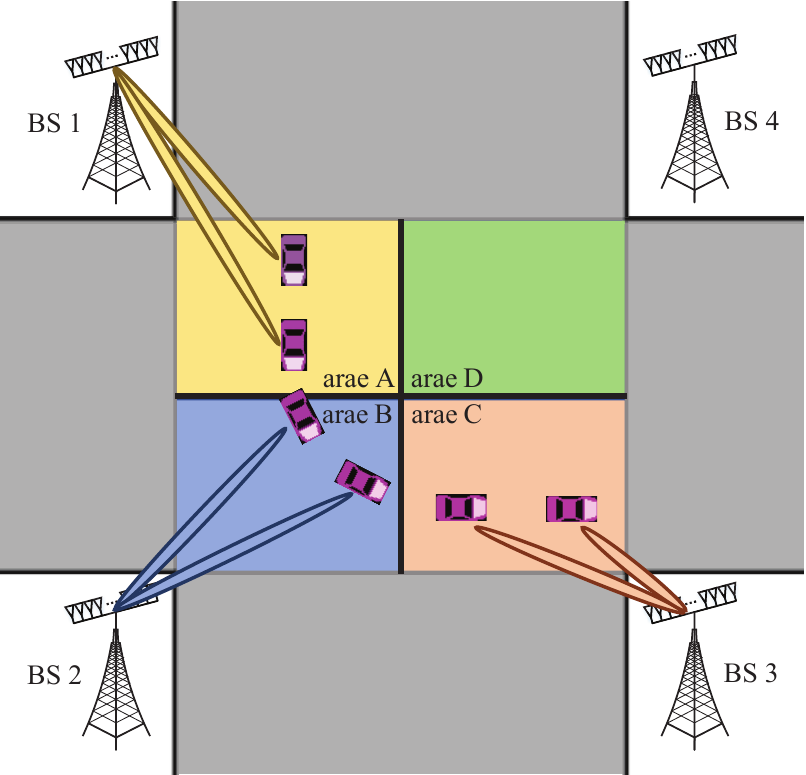}
    \caption{Area division based on user's distance from BS.}
    \label{area}
  \end{minipage}%
  \begin{minipage}[t]{0.5\linewidth}
    \centering
    \includegraphics[scale=0.8]{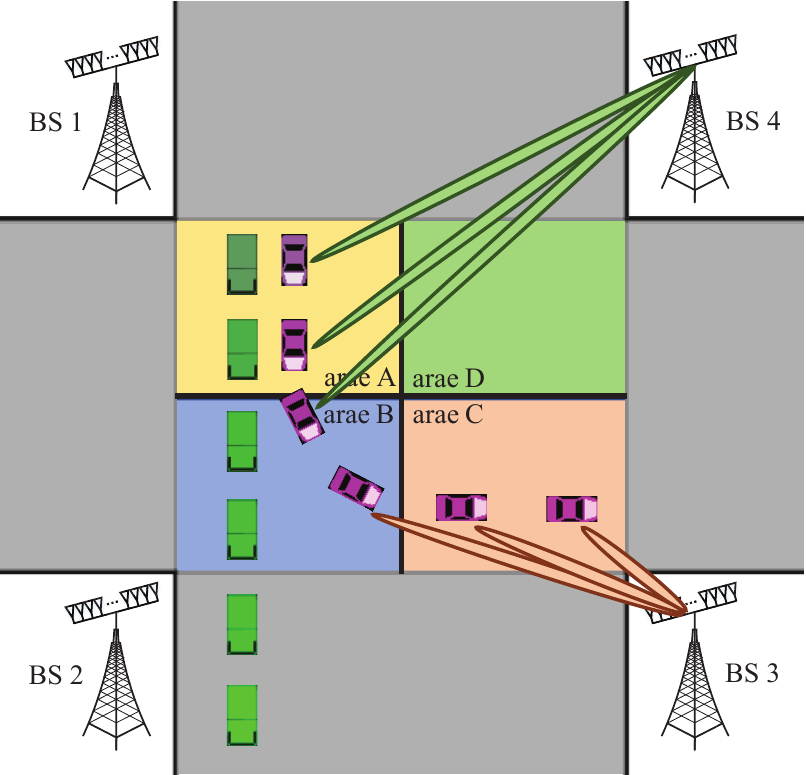}
    \caption{The optimal base station handover scheme when the vehicle is blocked.}
    \label{area-blocked}
  \end{minipage}
\end{figure}

We use the above LSTM structure to extract the features of each modality ($\mathbf{IS}^{1}_{i}$, $\mathbf{IS}^{2}_{i}$, $\mathbf{IS}^{3}_{i}$, $\mathbf{IS}^{4}_{i}, \mathbf{CS}_{i}$ and $\mathbf{LS}_{i}$) separately, defined as $\mathbf{IF}^{1}_{i}$ (image feature), $\mathbf{IF}^{2}_{i}$, $\mathbf{IF}^{3}_{i}$, $\mathbf{IF}^{4}_{i}, \mathbf{CF}_{i}$ (channel feature) and $\mathbf{LF}_{i}$ (location feature).
These features will be used for the main task and the auxiliary tasks.

\subsection{Multi-Task Learning Based BS Selection Module}\label{s_multi_task}
In this section, we propose the multi-task based BS selection module (BSM) to select the optimal BS in the next frame using the features extracted from the FEM.

For the mmWave communication system, if the environment is open and free of obstacles, the BS closest to the user has the strongest signal power.
In this case, the nearest BS is the optimal BS for the user.
Let us divide the crossroad into four areas according to the distance from the BS as shown in the Fig. \ref{area}.
Hence, the optimal BS of the user in the future has a strong correlation with the area where the user is located in the future.
We choose the task of predicting the area of the user in the next frame as the auxiliary task of selecting the optimal BS.

In Fig. \ref{area}. when the user is located in area A, BS 1 is the optimal BS to be connected, while when the user enters area B, BS 2 is more likely to be selected.
In Fig. \ref{area-blocked}, when the user is located in area A and is blocked by a large vehicle, BS 1 will no longer be the best communication base station, and the user needs to switch the BS in time to ensure smooth communication.
The blockage between the user and each BS affects the selection of the optimal BS.
Therefore, predicting the blockage is also very helpful for the accurate selection of the optimal BS.
We choose to predict the blockage in the next frame as another auxiliary task of selecting the optimal BS.

\begin{figure}[t]
\centering
\includegraphics[width=0.6\textwidth]{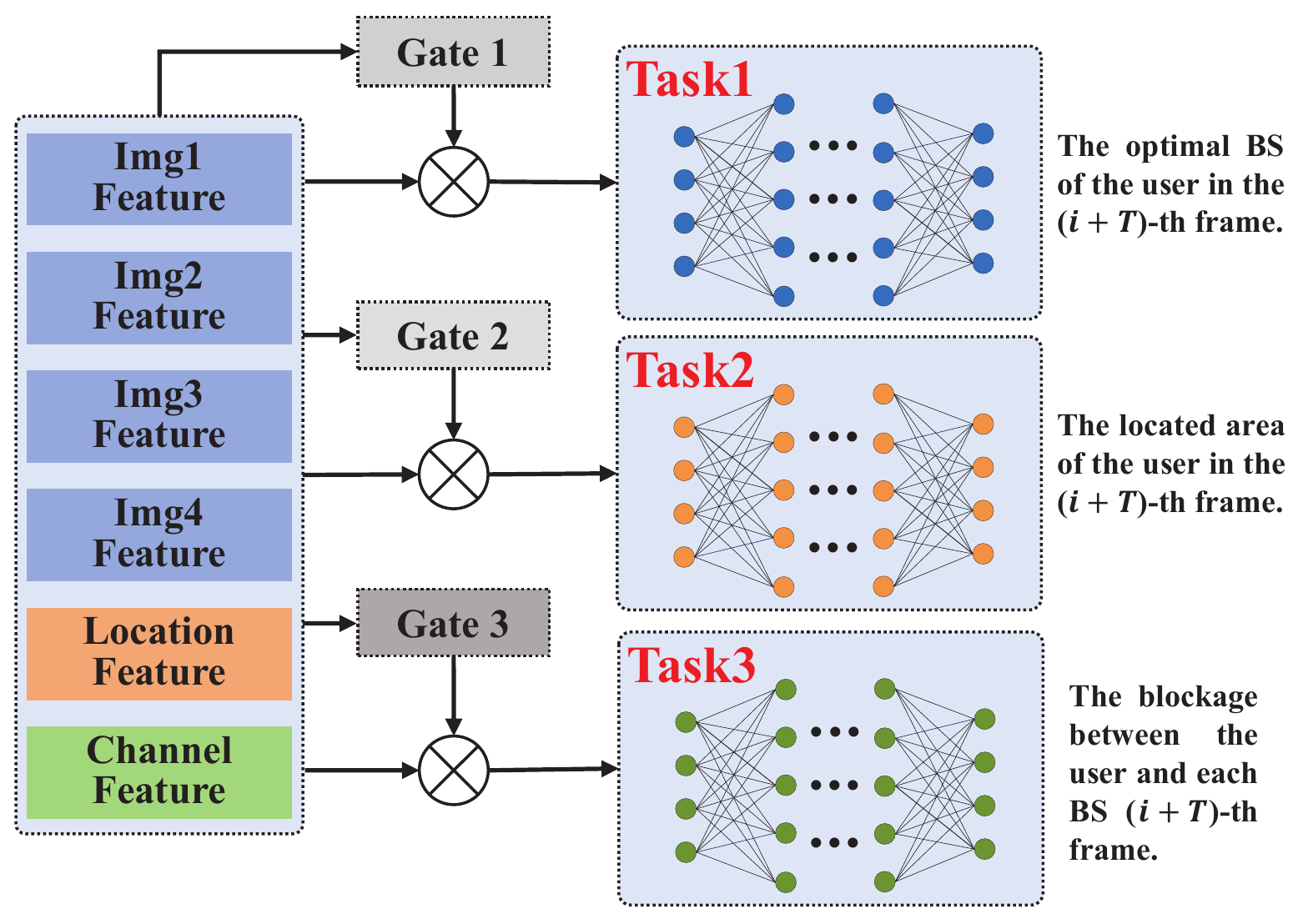}
\caption{Multi-task learning model for BS selection.}
\label{mtl_bs}
\end{figure}

The importance of the extracted features $\mathbf{IF}^{1}_{i}$, $\mathbf{IF}^{2}_{i}$, $\mathbf{IF}^{3}_{i}$, $\mathbf{IF}^{4}_{i}, \mathbf{CF}_{i}$ and $\mathbf{LF}_{i}$ varies across tasks.
For example, in regard to the task of selection the BS, channel, image and location information are all important, while for the task of predicting blockage, the image information may be more important.
Therefore, for different tasks, information of different modalities needs to be fused in different combinations.
We generate feature combination schemes for different tasks as shown in Fig. \ref{mtl_bs}.
A gate controller balances the influence of each feature for one task by performing a linear transformation on the features of all modalities and outputs an influence parameter vector.
\begin{equation}\label{gate_control}
\begin{aligned}
\bm\alpha_1 &= \mathbf{W}_{gate1} \cdot \left[(\mathbf{IM}^{1}_{i})^{T},(\mathbf{IM}^{2}_{i})^{T},(\mathbf{IM}^{3}_{i})^{T},(\mathbf{IM}^{4}_{i})^{T},(\mathbf{IM}^{4}_{i})^{T},(\mathbf{LM}_{i})^{T},(\mathbf{CM}_{i})^{T}\right]^{T},\\
\bm\alpha_2 &= \mathbf{W}_{gate2} \cdot \left[(\mathbf{IM}^{1}_{i})^{T},(\mathbf{IM}^{2}_{i})^{T},(\mathbf{IM}^{3}_{i})^{T},(\mathbf{IM}^{4}_{i})^{T},(\mathbf{IM}^{4}_{i})^{T},(\mathbf{LM}_{i})^{T},(\mathbf{CM}_{i})^{T}\right]^{T},\\
\bm\alpha_3 &= \mathbf{W}_{gate3} \cdot \left[(\mathbf{IM}^{1}_{i})^{T},(\mathbf{IM}^{2}_{i})^{T},(\mathbf{IM}^{3}_{i})^{T},(\mathbf{IM}^{4}_{i})^{T},(\mathbf{IM}^{4}_{i})^{T},(\mathbf{LM}_{i})^{T},(\mathbf{CM}_{i})^{T}\right]^{T},\\
\end{aligned}
\end{equation}
where $\mathbf{W}_{gate1},\mathbf{W}_{gate2},\mathbf{W}_{gate3}\in \mathbb{R}^{(4n_{img}+n_{cha}+n_{loc})\times 6}$ is the linear mapping matrix of the gate controller for each task, $\bm\alpha_k=[\alpha_{k,1},\alpha_{k,2},\alpha_{k,3},\alpha_{k,4},\alpha_{k,5},\alpha_{k,6}]$ is the influence parameter vector of each task, and $\alpha_{k,m}$ is the importance of the information of the $m$-th modality to the $k$-th \footnote{k=1 corresponds to the main task, k=2 corresponds to the subtask of the prediction area, and k=3 corresponds to the subtask of the prediction blockage.} task.
The input of each task $k$ is $\bm X_k=\left[(\mathbf{IM}^{1}_{i})^{T},(\mathbf{IM}^{2}_{i})^{T},(\mathbf{IM}^{3}_{i})^{T},(\mathbf{IM}^{4}_{i})^{T},(\mathbf{LM}_{i})^{T},(\mathbf{CM}_{i})^{T}\right]\cdot \bm\alpha_k$.
The network of each task consists of $L_k$ fully connected layers, and outputs
\begin{equation}
     \bm Y_k = NET_k(\bm{X}_k,\mathbf{\Theta}_k)  = \varphi_k^{(L_k)}(\cdots \varphi_k^{(1)}(\bm{X}_k))
    \label{sub6_info}
\end{equation}
where $\mathbf{\Theta}_k=\{{\mathbf{W}_k,\mathbf{b}_k}\}$ denotes the weights and biased of the fully connected layers, $\bm Y_k=\left[y_{k,1}, y_{k,2}, \cdots, y_{k,n_k}\right]^{T}$, and $n_k$ is the number of neurons in the $L_k$ layer. Moreover, $\varphi_k^{(l)}(\cdot)$ represents the non-linear function of the $l$-th layer for the $k$-th task and can be written as
\begin{equation}
    \varphi_k^{(l)}(\mathbf{x}_k) = \textup{ReLu}(W^{(l)}*\varphi_k^{(l-1)}(\mathbf{x}_k)+b^{(l)}) \quad l = 1,2,\cdots,L_k.
\end{equation}

Since task1 is a multi-class classification problem \cite{aly2005survey,bishop2006pattern}, the output layer of task1 is the softmax function
\begin{equation}\label{softmax}
\begin{aligned}
p_{1,i}=\text{softmax}\left(\bm{Y}_1\right)_{i}=\frac{\text{exp}\left(y_{1,i}\right)}{\sum\limits_{j=1}^{n_{1}}\text{exp}\left(y_{1,j}\right)},
\end{aligned}
\end{equation}
where $p_{1,i}$ represents the probability of the $i$-th BS being the optimal BS.
The loss function of task1 is the cross-entry loss
\begin{equation}\label{loss1}
\begin{aligned}
Loss_1=-\sum_{i=1}^{4}q_{1,i}\textup{log}(p_{1,i}),
\end{aligned}
\end{equation}
where $\bm q_{1} = [q_{1,1}, q_{1,2}, q_{1,3}, q_{1,4}]^{T}$ is the label of task1 and is a one-hot vector.
If the $i$-th BS is the optimal BS, then $q_{1,i}=1$; otherwise $q_{1,i}=0$.

Similarly, the output of task2 is
\begin{equation}\label{softmax-2}
\begin{aligned}
p_{2,i}=\text{softmax}\left(\bm{Y}_2\right)_{i}=\frac{\text{exp}\left(y_{2,i}\right)}{\sum\limits_{j=1}^{n_{2}}\text{exp}\left(y_{2,j}\right)},
\end{aligned}
\end{equation}
where $p_{2,i}$ represents the probability that the user will be located in the $i$-th area.
The loss function of task2 is the cross-entry loss
\begin{equation}\label{loss2}
\begin{aligned}
Loss_2=-\sum_{i=1}^{4}q_{2,i}\textup{log}(p_{2,i}),
\end{aligned}
\end{equation}
where $\bm q_{2} = [q_{2,1}, q_{2,2}, q_{2,3}, q_{2,4}]^{T}$ is the label task2 and is a one-hot vector.
If the user is located in the $i$-th area, then $q_{2,i}=1$; otherwise $q_{2,i}=0$.

Task3 predicts the blockage between the user and each BS in the ($i+T$)-th frame.
Denote the output as $\bm p_{3} = [p_{3,1}, p_{3,2}, p_{3,3}, p_{3,4}]^{T}$, and the label as $\bm q_{3} = [q_{3,1}, q_{3,2}, q_{3,3}, q_{3,4}]^{T}$.
Element $q_{3,i}$ indicates whether there is blockage between the user and the $i$-th BS.
If there is blockage between the user and the $i$-th BS, $q_{3,i}=0$; otherwise $q_{3,i}=1$.
The blockage between the user and each BS is independent of each other, and thus task3 is a multi-label classification problem \cite{tsoumakas2007multi}, and any element of $\bm q_{3}$ can be 0 or 1.
The activation function of the output layer is the sigmoid function
\begin{equation}\label{sigmoid}
\begin{aligned}
p_{3,i}=\text{Sigmoid}(y_{3,i})=\frac{1}{1+e^{-y_{3,i}}},
\end{aligned}
\end{equation}
where $p_{3,i}\in (0,1)$ represents the probability that the user and the $i$-th BS are not blocked, while $(1-p_{3,i})$ represents the probability that the user and the $i$-th BS are blocked.
The prediction for each $q_{3,i}$ is a binary classification problem \cite{kumari2017machine}.
The evaluation of the blockage prediction for \textbf{each} BS employs a cross-entropy loss.
Combining the predictions of the four BSs, the loss function of task3 is the Binary Cross Entropy (BCELoss) function
\begin{equation}\label{loss3}
\begin{aligned}
Loss_3=-\sum_{i=1}^{4}\left[q_{3,i}\textup{log}(p_{3,i})+(1-q_{3,i})\textup{log}(1-p_{3,i})\right].
\end{aligned}
\end{equation}

The three tasks share the underlying features and simultaneously predict the optimal BS, user's area, and user's blockage.

\subsection{Prior Knowledge Based Fine Tuning Module}
Due to the fast attenuation and poor diffraction ability of mmWaves, once there is blockage between the user and the $i$-th BS, the received power will attenuate exponentially.
Then the $i$-th BS cannot be the optimal one.
Using this common sense of communication, we design a prior knowledge based fine tuning network (FTM) to further increase the prediction accuracy of the optimal BS.
The input of FTM is $\bm{p}_{in}=\left[p_{in_{1}},p_{in_{2}},p_{in_{3}},p_{in_{4}}\right]=\bm p_{3}\odot \bm p_{1}$.
If there is blockage between the user and the $i$-th BS, then the output $p_{3,i}$ of task3 will be correspondingly close to 0.
\begin{figure}[t]
\centering
\includegraphics[width=0.7\textwidth]{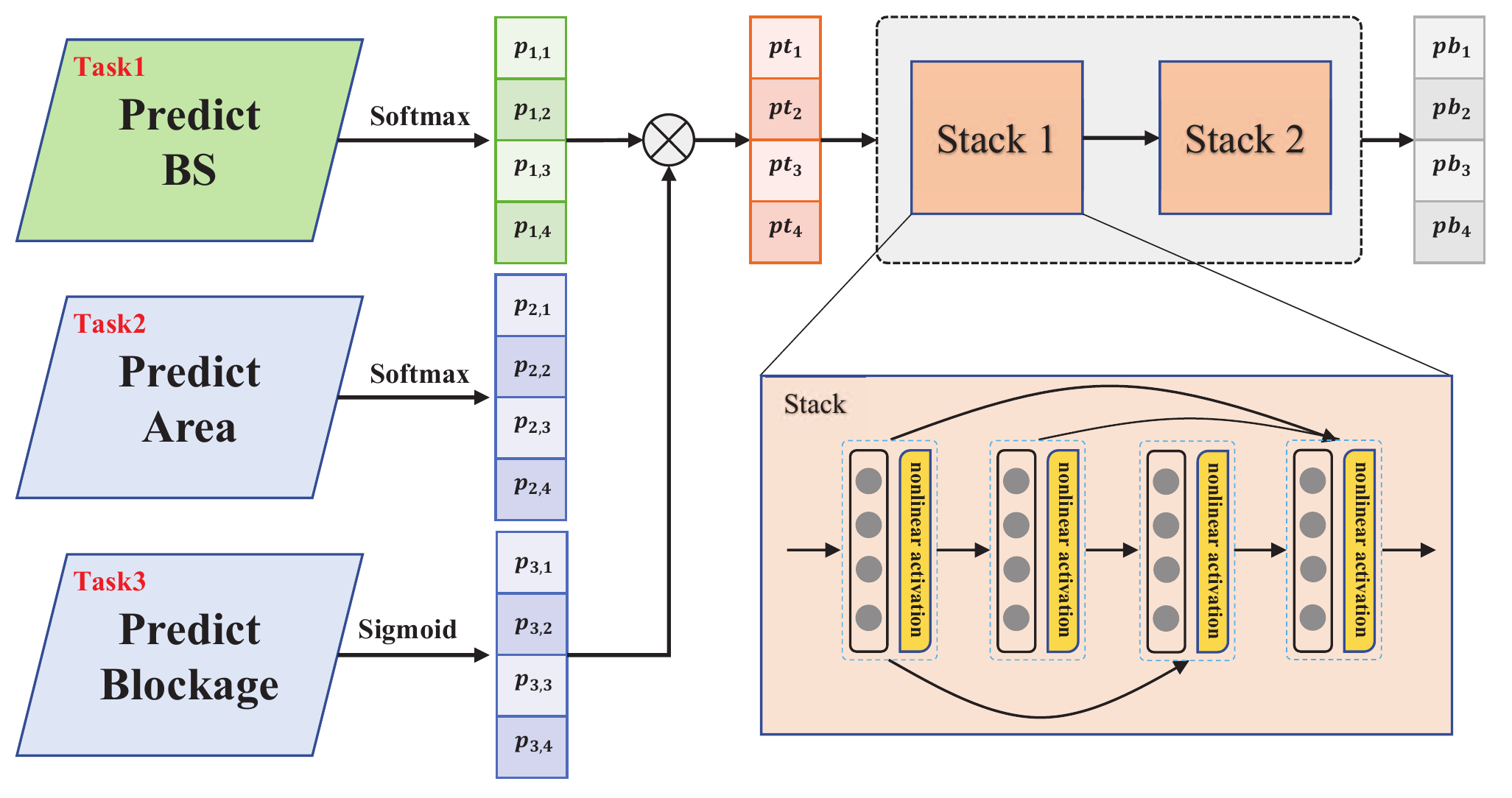}
\caption{Prior Knowledge Based Fine Tuning Module.}
\label{FTM}
\end{figure}
Therefore, the parameter $p_{in_{i}}$ will be further reduced.

The FTM consists of two stacks as shown in Fig. \ref{FTM}.
Each stack contains four fully connected layers, and the output of the $i$-th stack can be expressed recursively as
\begin{equation}\label{PKN-BLOCK}
\begin{aligned}
&\bm{Y}_{i,1}^{FTM}=\textup{ReLu}\left(\mathbf{W}^{(FTM)}_{i,1}*\bm{p}_{in}+\mathbf{bias}^{(FTM)}_{i,1}\right)\\
&\bm{Y}_{i,2}^{FTM}=\textup{ReLu}\left[\mathbf{W}^{(FTM)}_{i,2}*\left(\bm{Y}_{i,1}^{FTM}+\bm{p}_{in}\right)+\mathbf{bias}^{(FTM)}_{i,2}\right]\\
&\bm{Y}_{i,3}^{FTM}=\textup{ReLu}\left[\mathbf{W}^{(FTM)}_{i,3}*\left(\bm{Y}_{i,2}^{FTM}+\bm{Y}_{i,1}^{FTM}+\bm{p}_{in}\right)+\mathbf{bias}^{(FTM)}_{i,3}\right]\\
&\bm{Y}_{i,4}^{FTM}=\textup{ReLu}\left[\mathbf{W}^{(FTM)}_{i,4}*\left(\bm{Y}_{i,3}^{FTM}+\bm{Y}_{i,2}^{FTM}+\bm{Y}_{i,1}^{FTM}+\bm{p}_{in}\right)+\mathbf{bias}^{(FTM)}_{i,4}\right],
\end{aligned}
\end{equation}
In a stack, each layer is connected to each other and the input of each layer in the stack is the output of all previous layers.
This structure is called as densenet \cite{huang2017densely} that can well alleviate the gradient disappearance problem when the network is deep, and enhance feature propagation through feature reuse.
Through two stacks and a softmax layer, the final output of the network is
\begin{equation}\label{softmax-final}
\begin{aligned}
\bm{p}_{out}=\text{softmax}\left(\bm{Y}_{2,4}^{FTM}\right),
\end{aligned}
\end{equation}
where $\bm{p}_{out}=\left[p_{out_{1}},p_{out_{2}},p_{out_{3}},p_{out_{4}}\right]$ is the fine tuned probability vector. The loss function of the final output is the cross-entry loss
\begin{equation}\label{loss4}
\begin{aligned}
Loss_4=-\sum_{i=1}^{4}q_{4,i}\textup{log}(p_{out_{i}}),
\end{aligned}
\end{equation}
where $\bm q_{4}=\bm q_{2}$ is the label of the optimal BS.

Combining the loss of task1 (\ref{loss1}), task2 (\ref{loss2}), task3 (\ref{loss3}), and FTM (\ref{loss4}), the loss of the PBPN is
\begin{equation}\label{loss_all}
\begin{aligned}
Loss_{PBPN}=\sigma_1\cdot Loss_1 + \sigma_2\cdot Loss_2 + \sigma_3\cdot Loss_3 + \sigma_4\cdot Loss_4,
\end{aligned}
\end{equation}
where $\sigma_1, \sigma_2, \sigma_3, \sigma_4 \geq 0$ are the tuning parameters of the four losses.
\begin{figure}[t]
\centering
\includegraphics[width=0.78\textwidth]{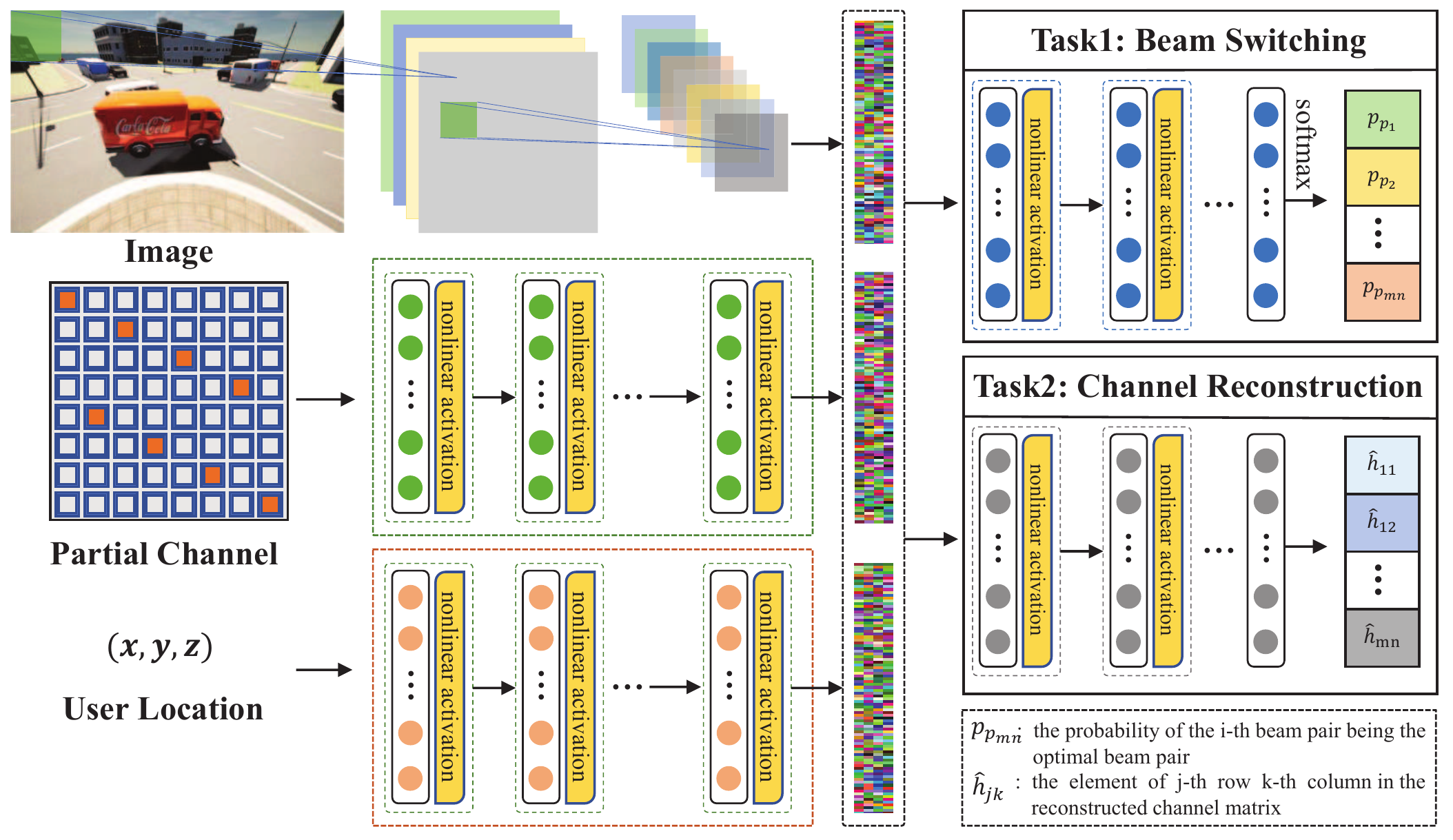}
\caption{Beam Pair Switching Network.}
\label{BPSN}
\end{figure}

\section{Beam Pair Switching Network}\label{s_predict2}
Once the PBSN selects the optimal BS $\hat{b}^{opt}$ of the ($i+T$)-th frame, the subsequent signal will be transmitted by BS $\hat{b}^{opt}$.
In this case, all possible beam pairs are $\mathcal F_{pair}=\left\{(\bm f_{t,1},\bm f_{r,1}),\dots,(\bm f_{t,N_t},\bm f_{r,N_r})\right\}$, where $\left|\mathcal F_{pair}\right|=N_t\cdot N_r$.
We then propose an image based beam pair switching network (BPSN).
The BPSN uses the image, channel, and position information in the ($i+T$)-th frame to switch the optimal beam pair.
According to (\ref{rate}), the optimal beam pair is directly related to the channel matrix of the user.
Therefore, learning how to estimate the full channel matrix of the user is helpful to predict the optimal beam pair.
We choose the task of reconstructing the complete channel matrix as the subtask.
The structure of BPSN is shown in Fig. \ref{BPSN}.

For the task of switching the optimal beam pair, the output of BPSN is the probability vector $\bm {p}_{p}=\left[p_{p_1},p_{p_2},\cdots,p_{p_{N_t\cdot N_r}}\right]$, where $p_{p_i}$ represents the probability of the $i$-th beam pair being the optimal beam pair.
The loss function of the task is the cross-entry loss
\begin{equation}\label{loss-1}
\begin{aligned}
Loss^{'}_{1}=-\sum_{i=1}^{N_t\cdot N_r}p_{l_i}\textup{log}(p_{p_i}),
\end{aligned}
\end{equation}
where $\bm{p}_l = \left[p_{l_1},p_{l_2},\cdots,p_{l_{N_t\cdot N_r}}\right]\in \{0,1\}^{N_t\cdot N_r}$ is a one-hot vector, and is the label of the beam pari.
Specifically, if the $i$-th beam pair in $\mathcal{F}_{pair}$ is the optimal beam pair, then there are $p_{l_i}=1, p_{l_j}=0 (\forall j\neq i)$.

For the subtask of reconstructing the complete channel matrix, the output of BPSN is the predicted channel matrix $\hat{\mathbf H}$, and the loss function of the task is the Normalized Mean Square Error (NMSE) loss function
\begin{equation}\label{loss-2}
\begin{aligned}
Loss^{'}_{2} = \frac{\|\mathbf{H}-\hat{\mathbf{H}}\|^{2}}{\|\mathbf{H}\|^{2}}.
\end{aligned}
\end{equation}

Combining the loss of the two tasks, the loss of the BPSN is
\begin{equation}\label{loss-all}
\begin{aligned}
Loss_{BPPN}=\sigma^{'}_1\cdot Loss^{'}_1 + \sigma^{'}_2\cdot Loss^{'}_2,
\end{aligned}
\end{equation}
where $\sigma^{'}_1$ and $\sigma^{'}_2 \geq 0$ are the tuning parameters of the two losses.

\section{Simulation Results}\label{simulation}
In this section, we  generate the dataset and evaluate the performance of the proposed BS selection network and beam pair switching network.
\subsection{Dataset Generation}
We consider a communication scenario at a crossroad, four BSs are located at the four corners, and the users are vehicles in the crossroad.
We used Carla \cite{dosovitskiy2017carla} to build this scenario, and capture visual images from Carla.\footnote{Carla has been developed from the ground up to support development, training, and validation of autonomous driving systems. In addition to open-source code and protocols, Carla provides open digital assets (urban layouts, buildings, vehicles) that were created for this purpose and can be used freely. Carla supports flexible specification of sensor suites, environmental conditions, full control of all static and dynamic actors, maps generation and much more.}
We equip an RGB camera at each BS and rotate it at specific angles such that the camera is facing the crossroad.
The parameters of the four cameras are shown in Table \ref{table1}.

\begin{figure}[t]
\centering
\includegraphics[width=0.35\textwidth]{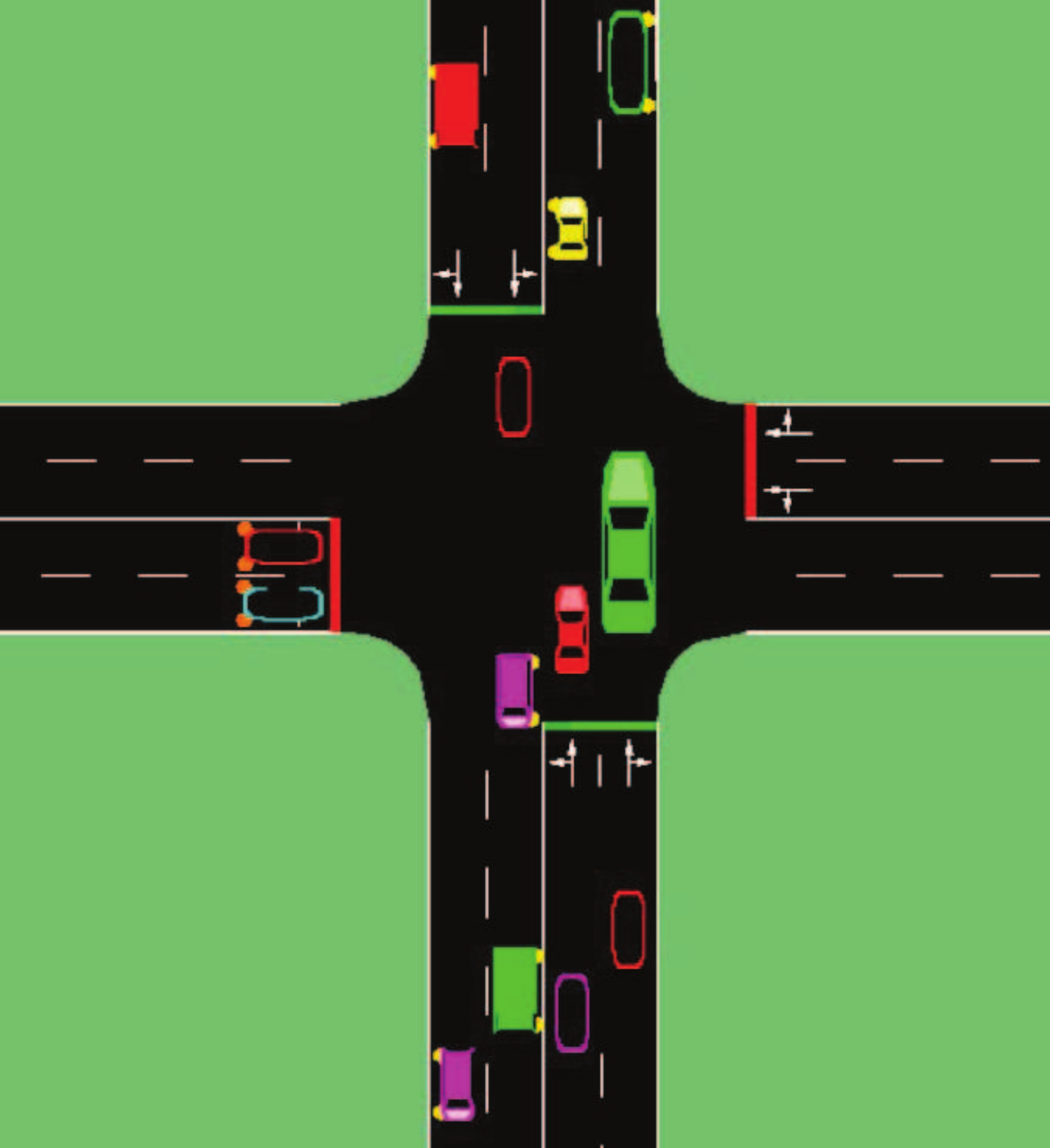}
\caption{The Traffic Flow in The Crossroad at A Certain Frame.}
\label{sumo}
\end{figure}

We use SUMO \cite{SUMO2018} software to design the traffic flow.\footnote{SUMO is a free and open source traffic simulation suite. It is available since 2001 and allows modelling of intermodal traffic systems, including road vehicles, public transport and pedestrians.}
We design the driving routes of many different vehicles where each vehicle follows the designated route from the starting point into the map, pass or not pass through the crossroad, and then disappear from the end point.
To ensure the rationality of the scenario and the diversity of the vehicle distribution at the crossroad, each vehicle may be assigned a different route.
Moreover, different kinds of vehicles will have different accelerations, decelerations and maximum speeds.
The traffic flow in the crossroad at a certain moment is shown in the Fig. \ref{sumo}.
During the simulation, the information about all vehicles on the map (not just those in crossroads) in each frame is recorded in a .xml file.
The recorded vehicle information includes the ID of the vehicle, the type of the vehicle, the location, the speed, and the serial number of the road where the vehicle is located.

\begin{table*}[t]
\renewcommand\arraystretch{}
\caption{Camera Parameters}\label{table1}
\centering
\begin{tabular}{p{1.5cm}<{\centering}|p{2cm}<{\centering}|p{2cm}<{\centering}|p{2cm}<{\centering}|p{1.5cm}<{\centering}|p{1.5cm}<{\centering}|p{1.5cm}<{\centering}}
\hline
\hline
Camera & Location-x & Location-y & Location-z & Pitch & Yaw & Roll \\
\hline
Camera1 & 8.7 & 10.8 & 4 & -30 & -45 & 0 \\
\hline
Camera2 & 8.7 & -11.1 & 4 & -30 & 45  & 0 \\
\hline
Camera3 & 28.2 & 10.8 & 4 & -30 & -135 & 0 \\
\hline
Camera4 & 28.2 & -11.1 & 4 & -30 & 135 & 0 \\
\hline
\hline
\end{tabular}
\end{table*}

Then we utilize the co-simulation feature developed by Carla to combine the scene picture in Carla with the traffic information in SUMO.
In each frame, Carla generates vehicles' models in the scene according to the vehicles' information provided by SUMO, and the four cameras capture images.
The vehicle information and corresponding images of each frame are saved and named after the frame number.
Vehicles in each frame are marked with bounding boxes.
At the same time, we calculate the located area of the vehicle according to the vehicle information, which is used for subsequent dataset combination.

\begin{figure}[t]
\centering
\includegraphics[width=0.45\textwidth]{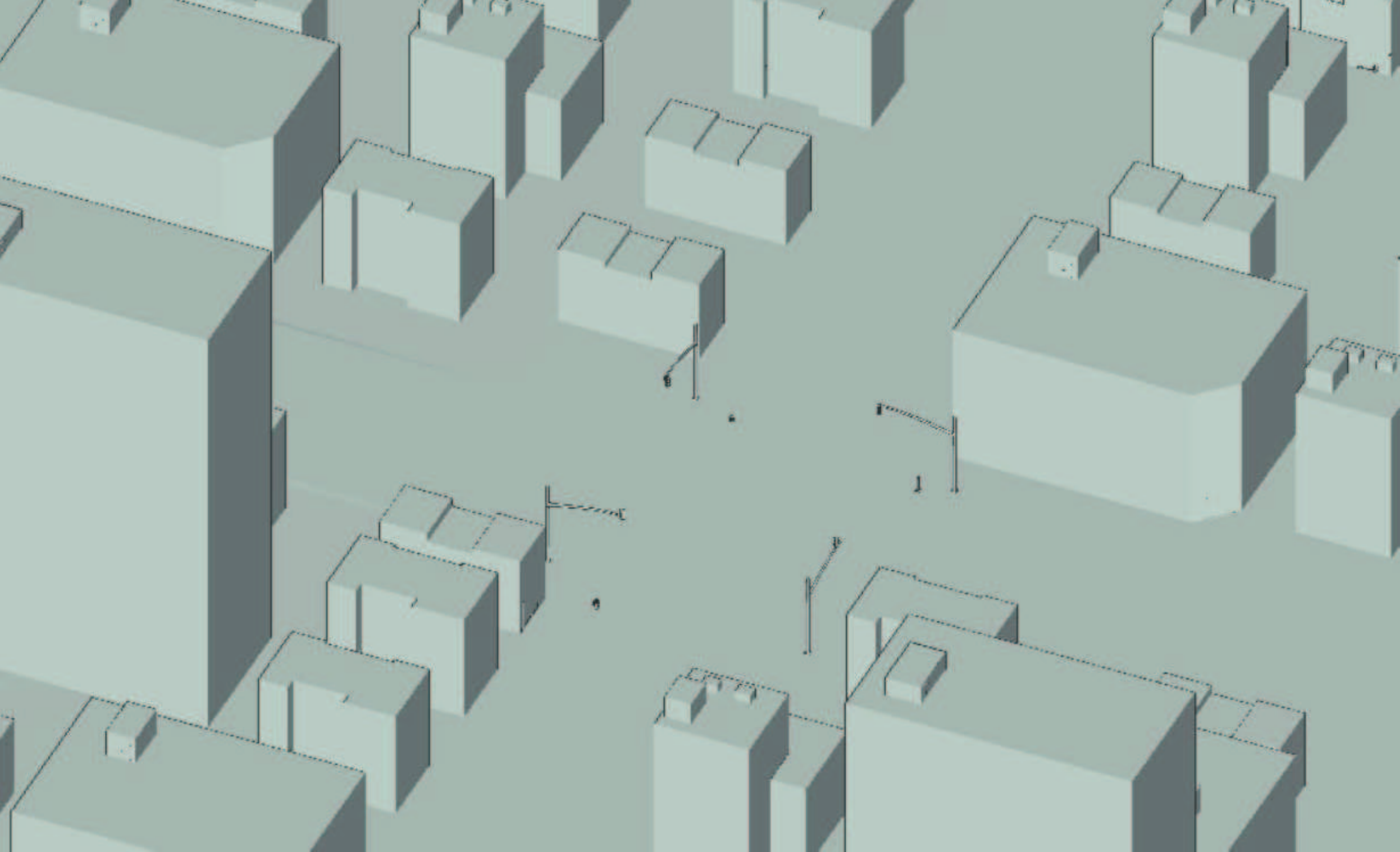}
\caption{The 3D model of the scenario in Wireless Insite.}
\label{WI}
\end{figure}

Next we use the 3D ray-tracing software Wireless InSite \cite{WinNT} to calculate the channel matrices of the users in each frame.\footnote{Wireless InSite is widely used in mmWave and massive MIMO research at both industry and academia.}
We first import the 3D model of the scenario, including the ground and buildings into the software as shown in Fig. \ref{WI}.
In each frame, we import all vehicles' model into Wireless Insite.\footnote{In order to simplify the modeling step, we ignore the detailed structure of the vehicle and use a cube with the same length, width and height to replace the vehicle model.}
Then, we set the material for the ground and buildings as concrete, and the material for the vehicles as metal.
BSs are placed at the four corners of the crossroad as the transmitters. For vehicles in crossroads, a set of antennas are equipped on the roof as the receivers.
We adopt the X3D ray model to simulate the parameters of the ray path between the transmitter and receiver.
The X3D ray model was developed by REMCOM company to provide a highly accurate, full 3D propagation model.\footnote{The simulation finally outputs the received power, received phase, time of arrival, angle of arrival and angle of departure.}
We select 25 paths with the strongest received power as effective paths to calculate the channel matrix according to (\ref{channelmodel}).
Then we calculate the optimal BS and corresponding beam pair of each vehicle according to (\ref{rate}).

According to the above steps, we obtain the images of the four cameras, the vehicles located in the crossroad, the bounding box, the channel matrix, the optimal BS, the optimal beam pair, and the located area of each vehicle for each frame.
For each vehicle inside the crossroad, we integrate the dataset according to the following strategy:
The camera images of $T$ consecutive frames, the user's location of $T$ consecutive frames, and the user's partial channel of $T$ consecutive frames are used as input datasets.
Among them, the user's location is calculated through the multi-view bounding boxes and the corresponding camera parameters.
The user's partial channel is the down-sampled result of the full channel matrix.
The user's area, the optimal BS and the beam pair in frame $T+1$ are taken as the label dataset.
\begin{table}[t]
\renewcommand\arraystretch{}
\caption{Camera Parameters}\label{table2}
\centering
\begin{tabular}{p{1.5cm}<{\centering}|p{2cm}<{\centering}|p{1.5cm}<{\centering}|p{2cm}<{\centering}}
\hline
\hline
$N_t$ & 64 & $N_r$ & 4 \\
\hline
Tx array & $8\times 8$ & Rx array & $4\times 1$ \\
\hline
$f$/GHz & 28 & $BW$/M & 400 \\
\hline
$T$ & 4 & Solver &  Adam \\
\hline
\multicolumn{2}{c|}{Dataset Size} &  \multicolumn{2}{c}{15000} \\
\hline
\multicolumn{2}{c|}{Dataset Split} &  \multicolumn{2}{c}{70\%-30\%} \\
\hline
\multicolumn{2}{c|}{Initial learning rate} &  \multicolumn{2}{c}{$1\times 10^{-3}$} \\
\hline
\hline
\end{tabular}
\end{table}

\subsection{Neural Network Training}
The configurations of the neural network for BS selection and beam pair switching are as follows:

1) Multi-Modal Feature Extraction Network for BS prediction:
The resolution of the image captured by the camera is $960\times640$ pixels.
To reduce the amount of computation, we downsample the image uniformly to $480\times320$ pixels.
The feature extractor for each frame of the four views is a three-layer convolutional neural network.
The (in\_channels, out\_channels, kernel\_size, stride) parameters of the first convolutional layer are ($3, 3, 13, 2$).
The parameters of the second and the third convolutional layer are ($3, 3, 9, 1$) and ($3, 3, 5, 1$).
The pool\_size of the three convolutional layers' maxpooling layer are ($2\times 2$).
The linear layer of the image feature extractor has 128 neurons.
We normalize the channel matrices over the entire dataset and downsample them by the factor of 8.
The location array is also normalized by the crossroads center.
The input size of the LSTM for the image feature sequence is 128.
The hidden size of that is 128. The number of layers is set to 1.
The input size of the LSTM for the channel sequence is 64.
The hidden size of that is 128. The number of layers is set to 1.
Moreover, the input size of the LSTM for the location sequence is 2.
The hidden size of that is 32. The number of layers is set to 1.

\begin{figure}[t]
\centering
\includegraphics[width=0.55\textwidth]{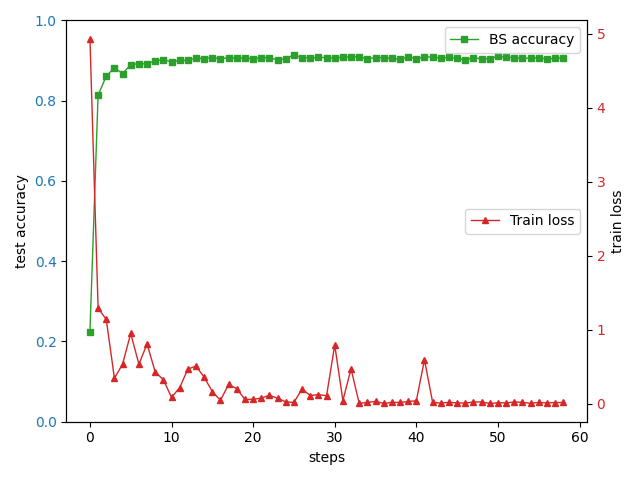}
\caption{The BS selection accuracy of the proposed PBSN.}
\label{BS_acc}
\end{figure}

2) Multi-Task BS Selection Network:
The dimension of the linear map $W_{gate1}, W_{gate2}, W_{gate3}$ are all $\mathbb{R}^{128\times 6}$.
The network of task1, task2, and task3 all consist of four linear layers, and the number of neurons in the four layer are 128, 64, 16, and 4, respectively.

3) Prior Knowledge Based Tuning Network:
Each layer of the Prior Knowledge Based Tuning Network contains four neurons.

4) Beam Pair Switching Network:
The image feature extractor includes three convolutional neural layers and one linear layer.
The (in\_channels, out\_channels, kernel\_size, stride, padding) parameters of the three layer are ($3, 3, 3, 1, 1$), ($3, 3, 5, 2, 2$) and ($3, 2, 5, 2, 1$) respectively.
The linear layer includes 64 neurons.
The channel extractor and location extractor both include three linear layers and the number of neurons are ($64, 64, 64$) and ($16, 32, 64$).
The beam pair prediction task and channel reconstruction  task both consist of three layers and the number of neurons are 256 and 512 respectively.

\begin{figure}[t]
\centering
\includegraphics[width=0.55\textwidth]{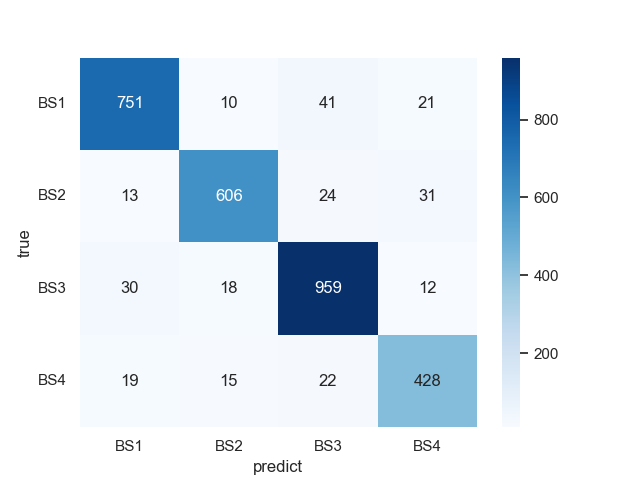}
\caption{The confusion matrix of BS selection.}
\label{confusion_matrix}
\end{figure}

\begin{figure}[t]
\centering
\includegraphics[width=0.55\textwidth]{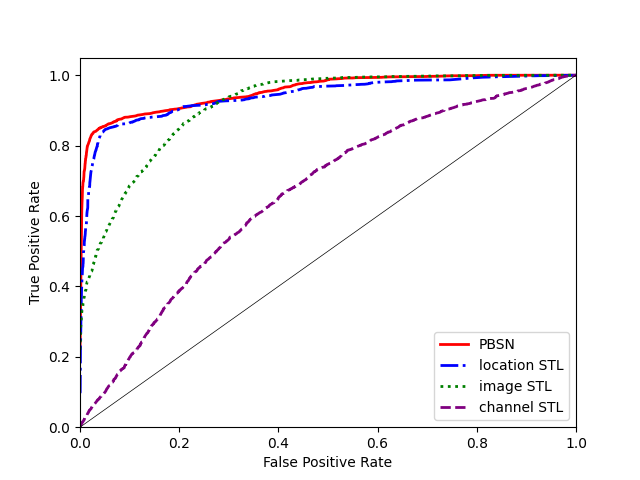}
\caption{The ROC of BS selection.}
\label{roc}
\end{figure}

\begin{figure}[t]
\centering
\includegraphics[width=0.55\textwidth]{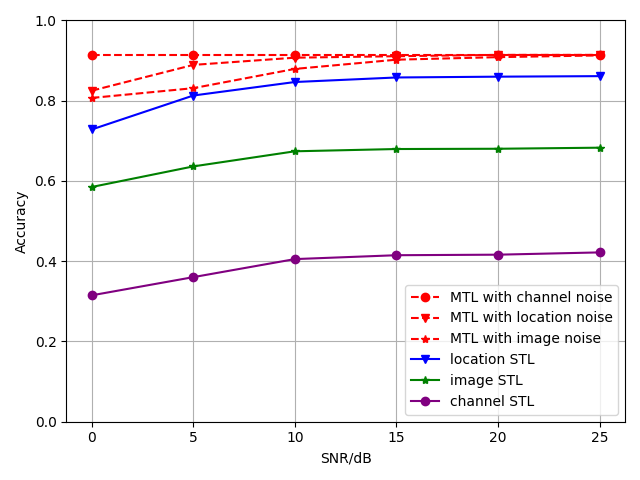}
\caption{The BS selection accuracy with noise.}
\label{BS_snr}
\end{figure}

\subsection{Performance Evaluation}

The BS prediction accuracy of PBSN is shown in Fig. \ref{BS_acc}.
The main axis plots the performance of the BS selection accuracy on the test set during the training progresses.
The secondary axis records the training loss during the training progresses.
The accuracy of BS prediction can reach 91.50\%.

The confusion matrix shows the classification of the network in each category. Specifically, Fig.\ref{confusion_matrix} plots the prediction results and true labels of each BS on the test dataset.
The number of correctly predicted results is shown on the main diagonal, and the various cases of incorrect predictions are shown on the non-main diagonal.
The precision of the four BSs are 92.37\%, 93.37\%, 91.68\%, 86.99\%, while the sensitivity (recall) of them are 91.25\%, 89.91\%, 94.11\%, 88.43\%.
Taking both the precision and sensitivity into consideration, the F1 Scores (2*precision*sensitivity/(precision+sensitivity)) of the BSs are 91.81\%, 91.61\%, 92.88\%,  87.75\%.

Then we compare the performance of PBSN with single modal single task learning (STL) models: location STL model, image STL model, and channel STL model.
The ROC curves\footnote{The receiver operating characteristic (ROC) curve, is a graphical plot that illustrates the diagnostic ability of a classifier. The larger the area under the ROC curve, the better the classifier.} of the PBSN and the single modal STL models are shown in Fig. \ref{roc}, where the proposed PBSN exhibits the best performance.
The location STL model is better than the image STL model while the image STL model is better than the channel STL model.

We next test the performance of the proposed PBSN under different noises.
The dotted lines draw the prediction accuracy of the PBSN under different modal noise, and the solid lines show the prediction accuracy of the location, the image and the channel single-modal STL model under the corresponding noise.
We can see that under the same noise level, the prediction accuracy of PBSN is consistently higher than that of the single-modal STL models.
Under the poor condition of SNR = 0dB, the prediction accuracy of PBSN can exceed 80\%.
Moreover, at 25dB SNR, the accuracy of PBSN can reach 91\%.

\begin{figure}[t]
\centering
\includegraphics[width=0.55\textwidth]{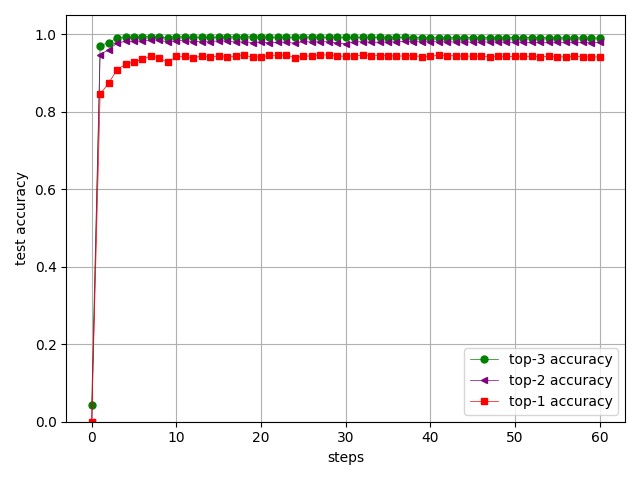}
\caption{The beam pair switching accuracy.}
\label{beam_acc}
\end{figure}

\begin{figure}[t]
\centering
\includegraphics[width=0.55\textwidth]{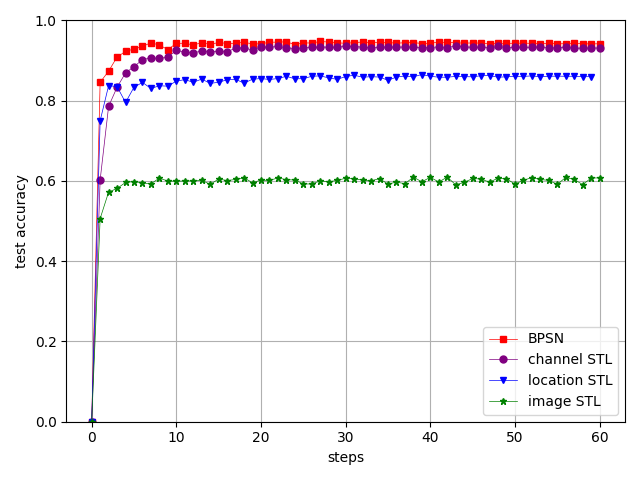}
\caption{The beam pair switching accuracy of different models.}
\label{beam_compare}
\end{figure}

After the BS in the $T+1$-th frame being selected, the objective BS switches the optimal beam pair using the proposed BPSN.
Taking BS 1 as an example, the prediction accuracy of beam pair is shown in the Fig. \ref{beam_acc}.
The top-1 accuracy is 94.47\%, the top-2 accuracy rate is greatly improved compared to the top-1 accuracy, which has increased to nearly 98.16\%.
Moreover, the top-3 accuracy rate reached 99.43\%.
In Fig. \ref{beam_compare}, we compare the proposed BPSN with three single modal STL models.
It is seen that the BPSN achieves the best accuracy.
Different from the results of BS selection, the accuracy of channel STL model is slightly lower than BPSN, and better than the location STL model.
The image STL model has the worst performance whose accuracy is only 60.45\%.

\begin{figure}[t]
\centering
\includegraphics[width=0.5\textwidth]{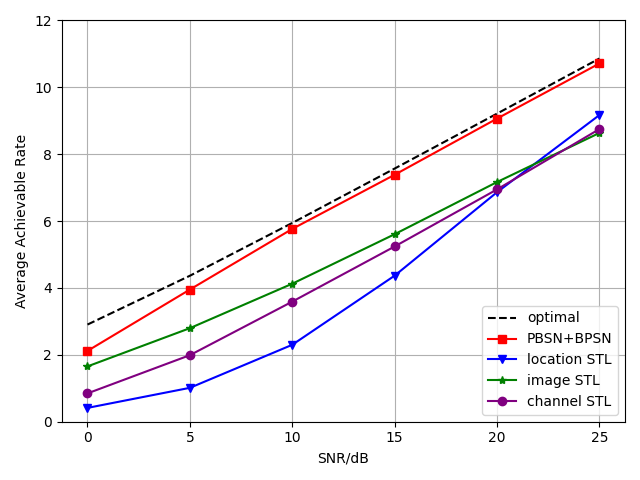}
\caption{The average achievable rate of different models.}
\label{rate_snr}
\end{figure}

Finally, combining the BS selection and beam pair switching, we test the average achievable rate of the proposed method.
The upper bound is the system rate at which the theoretically optimal BS is selected to communicate with the user and beamforming is performed using the optimal beam pair.
The achievable rate of the proposed `PBSN + BPSN' and the three single modal STL models are shown in Fig. \ref{rate_snr}.
The proposed `PBSN + BPSN' achieves a system rate close to the upper bound using only a small part of the channel matrix.
Moreover, the `PBSN + BPSN' is significantly better than the single modal STL model.
It can be seen from the above results that the proposed PBSN and BPSN can achieve near-optimal system rate with only a small part of the channel. Compared with the traditional exhaustive search algorithm, the pilot overhead for channel estimation is reduced by 7/8.
\section{Conclusions}\label{conclusion}

In this paper, we investigate the relationship between the scene information with the optimal BS and beam pair.
We propose a multi-camera view based proactive BS selection network that predicts the optimal BS in the $T+1$-th frame utilizing the scene information in the $0\sim T$ frames.
By using the multi-camera view images, the proposed PBSN can select the optimal BS in advance with a small part of channel.
The PBSN adopts the multi-task learning strategy that can improve the robustness of the network to noise.
The selecting accuracy of the BS achieves 91\% at SNR = 25dB.
Then we propose a beam pair switching network to predict the optimal beam pair of the objective BS in the $T+1$-th frame.
The switching accuracy of beam pair achieves 94.47\% while the top-3 accuracy is 99.43\%.
Moreover, the average achievable rate of the `PBSN + BPSN' is close to the upper bound and is significantly higher than the single modal single task learning model, while the pilot overhead for channel estimation is greatly reduced.

\begin{appendices}
\section{Calculation of 3D position}\label{appendix}
We first present the relationship between the world coordinates and the pixel coordinates.
Suppose the user's coordinate in the world coordinate system is $(X_w,Y_w,Z_w)$, and then its homogeneous coordinate\footnote{Homogeneous coordinates use N+1 dimensions to represent N-dimensional coordinates in order to deal with geometric problems in perspective space \cite{maxwell1952methods}. In perspective space, two parallel lines can meet at infinity. Using homogeneous coordinates, the translation of an object can be conveniently represented by a linear transformation.} is $(X_w,Y_w,Z_w,1)$.
The relative displacement of the camera and the user is recorded as $T=[t_x,t_y,t_z]^T$.
The rotation angles of the camera along each axis are $(\phi_x,\phi_y,\phi_z)$.
The rotation matrices along each axis are
\begin{equation}\label{rotation-matrices}
\begin{aligned}
R_x = \left[
\begin{matrix}
\textup{cos}\phi_x & -\textup{sin}\phi_x & 0 \\
\textup{sin}\phi_x & \textup{cos}\phi_x & 0 \\
0 & 0 & 1
\end{matrix}
\right],
R_y = \left[
\begin{matrix}
1 & 0 & 0 \\
0 & \textup{cos}\phi_y & \textup{sin}\phi_y \\
0 & -\textup{sin}\phi_y & \textup{cos}\phi_y
\end{matrix}
\right],
R_z = \left[
\begin{matrix}
\textup{cos}\phi_z & 0 & -\textup{sin}\phi_z \\
0 & 1 & 0 \\
\textup{sin}\phi_z & 0 & \textup{cos}\phi_z
\end{matrix}
\right].
\end{aligned}
\end{equation}
Hence the rotation matrix of the camera is $R = R_x\cdot R_y\cdot R_z$.
Then the coordinates of the user in the camera coordinate system are
\begin{equation}\label{camera-coordinate}
\left[
\begin{matrix}
X_c \\
Y_c \\
Z_c \\
1
\end{matrix}
\right]=
\left[
\begin{matrix}
R & T \\
0 & 1
\end{matrix}
\right]
\left[
\begin{matrix}
X_w \\
Y_w \\
Z_w \\
1
\end{matrix}
\right].
\end{equation}
According to the principle of camera imaging \cite{chuang2005camera}, we can calculate the coordinates $(x,y)$ of the user in the image coordinate system as
\begin{equation}\label{image-coordinate}
Z_c \left[
\begin{matrix}
x \\
y \\
1
\end{matrix}
\right]=
\left[
\begin{matrix}
f_x & 0 & 0 & 0 \\
0 & f_y & 0 & 0 \\
0 & 0 & 1 & 0
\end{matrix}
\right]
\left[
\begin{matrix}
X_c \\
Y_c \\
Z_c \\
1
\end{matrix}
\right],
\end{equation}
where $f_x$ and $f_y$ are the focal lengths of the camera lens on the x and y axes, respectively.
Next we convert the image coordinate to pixel coordinate.
Suppose one pixel occupies $dx$ unit lengths in the x-axis direction and $dy$ unit lengths in the y-axis direction.
The pixel coordinate of the center of the image is $(u_0,v_0)$.
Then the pixel coordinate of the user is $(u,v)$ that satisfies
\begin{equation}\label{pixel-coordinate}
\begin{aligned}
Z_c \left[
\begin{matrix}
u \\
v \\
1
\end{matrix}
\right]=
\left[
\begin{matrix}
\frac{1}{dx} & 0 & u_0 \\
0 & \frac{1}{dy} & v_0 \\
0 & 0 & 1
\end{matrix}
\right]
\left[
\begin{matrix}
x \\
y \\
1
\end{matrix}
\right],
\end{aligned}
\end{equation}
According to the properties of homogeneous coordinates, the relationship between the user's world coordinates and pixel coordinates can be expressed as
\begin{equation}\label{relationship-coordinate}
\left[
\begin{matrix}
u \\
v \\
1
\end{matrix}
\right]=
\left[
\begin{matrix}
\frac{1}{dx} & 0 & u_0 \\
0 & \frac{1}{dy} & v_0 \\
0 & 0 & 1
\end{matrix}
\right]
\left[
\begin{matrix}
f_x & 0 & 0 & 0 \\
0 & f_y & 0 & 0 \\
0 & 0 & 1 & 0
\end{matrix}
\right]
\left[
\begin{matrix}
R & T \\
0 & 1
\end{matrix}
\right]
\left[
\begin{matrix}
X_w \\
Y_w \\
Z_w \\
1
\end{matrix}
\right].
\end{equation}
\end{appendices}
We can get the 3D coordinates of the user by performing the inverse process of (\ref{relationship-coordinate}).
When the pixel coordinates of more than two views are obtained, the 3D coordinates of the user can be solved.

\linespread{1.02}

\bibliographystyle{IEEEtran}

\end{document}